\documentclass[12pt]{article}
\usepackage{a4wide}
\usepackage{amssymb,amsmath,amsthm,amscd}
\begin{document}
{\renewcommand{\thefootnote}{\fnsymbol{footnote}}
\hfill  IGC--08/10--2\\
\medskip
\begin{center}
{\LARGE  Dilaton Gravity, Poisson Sigma Models\\ and Loop Quantum Gravity}\\
\vspace{1.5em}
Martin Bojowald\footnote{e-mail address: {\tt bojowald@gravity.psu.edu}}
and Juan D.~Reyes\footnote{e-mail address: {\tt jdr234@psu.edu}}
\\
\vspace{0.5em}
Institute for Gravitation and the Cosmos,\\
The Pennsylvania State
University,\\
104 Davey Lab, University Park, PA 16802, USA\\
\vspace{1.5em}
\end{center}
}

\setcounter{footnote}{0}

\newcommand{\dd}{\delta}                       %define d operator in infinite-dim. phase space
\newcommand{\vd}[2]{\frac{\delta #1}{\delta #2}}   %define variational derivative   
\newcommand{\Ef}{E^\varphi}
\newcommand{\Kf}{K_\varphi}

\newcommand{\tG}{\tilde{C}}
\newcommand{\G}{C}

\newcommand*{\R}{{\mathbb R}}
\newcommand*{\N}{{\mathbb N}}
\newcommand*{\Z}{{\mathbb Z}}
\newcommand*{\Q}{{\mathbb Q}}
\newcommand*{\C}{{\mathbb C}}

\begin{abstract}
 Spherically symmetric gravity in Ashtekar variables coupled to
 Yang--Mills theory in two dimensions and its relation to dilaton
 gravity and Poisson sigma models are discussed. After introducing its
 loop quantization, quantum corrections for inverse triad components
 are shown to provide a consistent deformation without anomalies. The
 relation to Poisson sigma models provides a covariant action
 principle of the quantum corrected theory with effective
 couplings. Results are also used to provide loop quantizations of
 spherically symmetric models in arbitrary $D$ space-time dimensions.
\end{abstract}

\section{Introduction}

Approaches to quantum gravity imply characteristic corrections to
classical general relativity, by which their viability can be tested
and which may eventually give rise to observable indications. Distinct
approaches often lead to corrections of very different forms, but this
is not necessarily contradictory: Formulations are usually difficult to
compare directly which may shroud similar implications of
different-looking terms, especially if canonical formulations on the
one hand and Lagrangian ones on the other are used. In particular, it
is not always straightforward to derive action principles corresponding to
quantum corrected Hamiltonian formulations.

In the canonical approach underlying loop quantum gravity
\cite{Rov,ALRev,ThomasRev}, characteristic corrections in equations of
motion arise from three sources: corrections of inverse components of
the densitized triad which is used instead of a spatial metric
\cite{QSDI,InvScale}, higher order corrections of connection
components (due to the use of holonomies \cite{LoopRep}), and genuine
quantum effects due to the back-reaction of higher moments of a state
on its expectation values. The latter is generic for any interacting
quantum system, where it leads, e.g., to effective potentials; for a
canonical treatment see \cite{EffAc,Karpacz}. The first two sources of
corrections are characteristic of loop quantum gravity and directly
related to the underlying discreteness of its spatial geometry.

In homogeneous models of loop quantum cosmology \cite{LivRev}, these
corrections have been analyzed in quite some detail. In such settings,
however, consistency conditions (such as anomaly freedom) trivialize
which leaves much freedom in the form of corrections and a high level
of ambiguity in any effective action which might correspond to the
corrected equations of motion. In this article, we thus consider
spherically symmetric models \cite{SphSymm} as an intermediate step
where the consistency issue is non-trivial but calculations are still
manageable. As it turns out, correction terms in effective actions are
then determined much more uniquely.

We will exploit the fact that spherically symmetric gravity, or more
generally 2-dimensional dilaton gravity \cite{Strobl,DilatonRev}, can
be formulated as a Poisson sigma model (PSM) which algebraically has a
very rigid structure: a consistent deformation (such as an
anomaly-free quantization) of a PSM can only be another PSM
\cite{Izawa}. There is thus a clear way of interpreting any consistent
way of introducing quantum corrections. Applying this to loop quantum
gravity will then provide a covariant interpretation of corrections in
the form of a corrected Poisson structure of the target manifold, or
an effective dilaton potential. We also note that PSMs play a role as
models of string theory, such that an analysis along the lines
followed here may shed light on the relation between loop quantum
gravity and string theory.

The specific correction analyzed here is only the inverse triad term,
which is one of the terms directly linked to the discreteness, while,
for now, we ignore holonomy corrections and quantum back-reaction. We
will show that this can be implemented as a consistent deformation
under certain conditions on the phase-space dependence of the
correction, and then relate it to changes in the underlying Poisson
structure of the PSM formulation. To that end, we first review the
different underlying classical formulations in Sec.~\ref{s:2dGrav} and
derive an explicit canonical transformation to relate them in
Sec.~\ref{s:Relation}. After introducing quantum corrections at the
level of the loop formulation, as it is reviewed in Sec.~\ref{s:Loop}
in nearly self-contained form, we will translate this back to the PSM
formulation in Sec.~\ref{s:Consistent}. The analysis will be done in
the presence of a Yang--Mills source which can, as detailed in
Sec.~\ref{s:YM}, be added as an extension of the original PSM.

%%%%%%%%%%%%%%%%%%%%%%%%%%%%%%%%%%%%%%%%%%%%%%%
%%%% SUMMARY OF 2D GRAVITY AND PSM'S %%%%%%%%%%%
%%%%%%%%%%%%%%%%%%%%%%%%%%%%%%%%%%%%%%%%%%%%%%%

\section{Formulations of gravity in two dimensions}
\label{s:2dGrav}

In two dimensions the analog of the Einstein--Hilbert action in vacuum
is trivial, but the presence of extra fields gives rise to interesting
models. Such fields may, for instance, arise after dimensional
reduction of a field theory in four or higher dimensions.  In the case
of 2d gravity this leads to the presence of the dilaton field.  Most
of the studied gravity theories in two dimensions can be described by
the so-called Generalized Dilaton Gravity action
\begin{equation} \label{GDGaction}
S[\mathbf{g},\Phi]=\frac{1}{2}\int_M d^2x\sqrt{-\mathbf{g}}
\left(\mathbf{F}(\Phi)R(\mathbf{g})+\mathbf{U}(\Phi)\mathbf{g}^{\mu\nu}
\partial_\mu\Phi\partial_\nu\Phi + \mathbf{V}(\Phi)\right)
\end{equation} 
which is the most general diffeomorphism invariant action giving
second order differential equations for a metric $\mathbf{g}$ and a
scalar dilaton field $\Phi$ on a two-dimensional manifold $M$; for a
comprehensive review see \cite{DilatonRev}. Here, $\mathbf{F}$,
$\mathbf{U}$ and $\mathbf{V}$ are three functions parameterizing
different models (which should be sufficiently well behaved, and
$\mathbf{F}$ being invertible).

\subsection{Poisson Sigma Models}

Poisson sigma models are a more general and unifying structure
encompassing and generalising 2d gravity theories as well as 2d
Yang-Mills theories \cite{Ikeda,PSM,IkedaIzawa,Strobl}.
These are topological two-dimensional field theories which encode all
the content of a particular model in a single Poisson tensor
$\mathcal{P}$ defined on an $n$-dimensional target manifold $N$ (in
local coordinates $\mathbf{X}^i$,
$\mathcal{P}=\frac{1}{2}\mathcal{P}^{ij}(\mathbf{X})\frac{\partial}{\partial
\mathbf{X}^i}\wedge\frac{\partial}{\partial \mathbf{X}^j}$).

For a given Poisson tensor, the corresponding PSM action has the form
\begin{equation} \label{PSMaction}
S_{\rm PSM}=\int_M \mathbf{A}_i\wedge d\mathbf{X}^i+\frac{1}{2}\mathcal{P}^{ij}
\mathbf{A}_i \wedge \mathbf{A}_j
\end{equation}
or, written explicitly with coordinates $x^\mu$ on $M$,
\[
S_{\rm PSM}=\int_M dx^\mu\wedge dx^\nu\left[\mathbf{A}_{i\mu}(x)
\frac{\partial \mathbf{X}^i}{\partial x^\nu}(x) +
\frac{1}{2}\mathcal{P}^{ij}(\mathbf{X}(x))
\mathbf{A}_{i\mu}(x)\mathbf{A}_{j\nu}(x)\right]\,.
\]
The dynamical fields are $\mathbf{X}^i(x^\mu)$, which parameterize a map
$\mathcal{X}\colon M\to N$ from the two-dimensional spacetime manifold
$M$ to the target Poisson manifold $N$, as well as
$\mathbf{A}_i=\mathbf{A}_{i\mu}dx^\mu$, a one-form on $M$ taking
values in $\mathcal{X}^*(T^*N)$.

More abstractly, the pair $(\mathbf{X}^i,\mathbf{A}_i)$ defines a vector
bundle morphism $TM\to T^*N$ with base map $\mathcal{X}$, so that the
action may be viewed as a functional of vector bundle morphisms.  The
equations of motion of the PSM may be shown to require these morphisms
to preserve the standard Lie algebroid structures\footnote{A Lie
algrebroid is essentially a fiber bundle with a Lie bracket defined on
its sections as well as an anchor map from the bundle to the tangent
bundle over the same base manifold. A Lie algebra is a Lie algebroid
whose base manifold is a single point.} on $TM$ and $T^*N$:
solutions to PSMs are Lie algebroid morphisms, and gauge symmetries
are related to homotopies of morphisms \cite{LieAlgebroidPSM}.

The key step in establishing 2d gravity as a PSM is the reformulation
of (\ref{GDGaction}) in first order form by using Einstein-Cartan
variables: dyads and the spin connection instead of the metric.
First, using the field redefinition $\phi=\mathbf{F}(\Phi)$ and
replacing $\mathbf{U}$ by $\mathbf{U}/\mathbf{F}'\,^2$, the
coefficient $\mathbf{F}$ for the curvature term may always be assumed
to be the identity function. The kinetic term can be eliminated
by means of a conformal transformation $g:=\Omega^2(\phi)\mathbf{g}$,
with
\[
\Omega(\phi)=\exp\left(\int^\phi\frac{\mathbf{U}(z)}{2}dz + {\rm const}\right)
\]
and $V=-\mathbf{V}(\phi)/\Omega^2(\phi)$. The action
\[  
S=\frac{1}{2}\int_M d^2x\sqrt{-g}(\phi R - V(\phi))
\]
can then be expressed in first order form using dyads $e_\mu^adx^\mu$
and connection one-forms $\omega_\mu\,^a_bdx^\mu$:
\begin{equation}  \label{FOGaction}
S=-\int_M \phi d\omega +\frac{1}{2}V(\phi)\varepsilon + X_aDe^a\,.
\end{equation}
Here, we have used the two-dimensional identity
$R\varepsilon=-2d\omega$ where $\varepsilon$ is the two-dimensional
volume form and $\omega$ is defined by
$\omega^a_b=\omega\varepsilon^a_b$, with $\varepsilon^a_b$ being the
single generator of the Lorentz gauge algebra
$\mathfrak{so}(1,1)$. Lagrange multipliers $X^a$ are introduced to
enforce the condition of torsion freedom; see App.~\ref{altApproach}
for theories with torsion.

%Alternatively, it can be shown \cite{GrumillerKV} that
%(\ref{GDGaction}) is equivalent to a theory with non trivial torsion,
%characterized by the more general action
%\[
%S=-\int_M \phi d\omega -W(X_aX^a,\phi)\varepsilon + X_aDe^a
%\]
%with $W(X_aX^a,\phi)=\mathbf{U}(\phi)\frac{X_aX^a}{2}+\mathbf{V}(\phi)$. 

%%%%%%%%%%%%%%%%%%%%%%%%%%%%%%%%%%%%%%%%%%%%%%%%%%%
%%%%%%%   EoM for 2d gravity %%%%%%%%%%%%%%%%%%%%%%
%%%%%%%%%%%%%%%%%%%%%%%%%%%%%%%%%%%%%%%%%%%%%%%%%%%

\subsubsection{Equations of motion}
Variation of (\ref{FOGaction}) with respect to $\phi$, $\omega$, $X_a$
and $e^a$ respectively gives the equations of motion
\begin{gather}
d\omega+\frac{1}{2}V'(\phi)\varepsilon=0 \\
d\phi+X_a\varepsilon^a\,_b e^b=0  \label{EoM2} \\
De^a=de^a+\varepsilon^a\,_b\omega\wedge e^b=0  \label{TorsionFreeCond}\\
\frac{1}{2}V(\phi)\varepsilon_{ab}e^b+dX_a+\varepsilon_{ab}X^b\omega=0\,.
\end{gather}

It is convenient to introduce a light cone basis $e^\pm$ for dyads, so
that $g_{\mu\nu}=2e^+_{(\mu}e^-_{\nu)}$ and raising and lowering
indices is accomplished by replacing a lower $+$ ($-$) by an upper $-$
($+$) and vice versa.  Solving the condition (\ref{TorsionFreeCond})
for torsion freedom with coordinates $(t,x)$ on $M$ and a light cone
basis, we get the spin connection in terms of the dyad:
\begin{align}
\omega_x&=\frac{ (e_x^-e_x^+)^\cdot - (e_t^-e_x^++e_t^+e_x^-)' + 
e_t^-e_x^+\,' + e_t^+e_x^-\,' }{e_t^+e_x^--e_x^+e_t^-} \label{spinConn} \\
\omega_t&=\frac{\dot{e}_x^-e_t^++\dot{e}_x^+e_t^--(e_t^-e_t^+)'}{e_t^+e_x^-
-e_x^+e_t^-} 
\end{align}
From Eq.~(\ref{EoM2}) we obtain
\begin{equation}
X^\mp=\frac{\phi'e_t^\mp-\dot{\phi}e_x^\mp}{e_t^+e_x^--e_x^+e_t^-}  \label{Xpm}\,.
\end{equation}
These equations will be useful below.

%%%%%%%%%%%%%%%%%%%%%%%%%%%%%%%%%%%%%%%%%%%%%%%%%%%%%%%%%
%%%%%%% SPHERICALLY SYMMETRIC GRAVITY AS A PSM %%%%%%%%%%
%%%%%%%%%%%%%%%%%%%%%%%%%%%%%%%%%%%%%%%%%%%%%%%%%%%%%%%%%

\subsubsection{Spherically Symmetric Gravity as a Poisson sigma model}

To relate dilaton gravity and PSMs to spherically reduced gravity in
four dimensions, we start from the Einstein-Hilbert action
\[
S_{\rm EH}=\frac{1}{16\pi G}\int_{M\times S^{D-2}}d^Dx\;\sqrt{-\,^Dg}\;^DR
\]
in $D$ dimensions and insert the ansatz 
\begin{equation}
ds^2=\mathbf{g}_{\mu\nu}(x^\mu)dx^\mu dx^\nu +\Phi^2(x^\mu)d\Omega^2_{S^{D-2}}  
\label{SSmetric}
\end{equation}
for a spherically symmetric metric with $\mathbf{g}_{\mu\nu}$ a metric
of signature $(-,+)$ on the two-dimensional spacetime $M$, and
$d\Omega^2_{S^{D-2}}$ the area element of the $(D-2)$-sphere $S^{D-2}$
(for the 2-sphere $d\Omega^2=d\vartheta^2+\sin^2\vartheta
d\varphi^2$). After integration of the angular variables (see e.g.\
App.~B of \cite{GroupArea} or App.~C of \cite{GrumillerThesis}), the
reduced 2d dilaton action is
\begin{equation}
S=\frac{\mathcal{O}_{D-2}}{16\pi G}\int_M
d^2x\sqrt{-\mathbf{g}}\left[\Phi^{D-2}
\mathbf{R}(\mathbf{g})+(D-2)(D-3)\Phi^{D-4}\mathbf{g}^{\mu\nu}\partial_\mu\Phi\partial_\nu\Phi+(D-2)(D-3)\Phi^{D-4}\right]\,.
\label{SSreducedAction}
\end{equation}
with $\mathcal{O}_{D-2}$ the volume of $S^{D-2}$.
 
Defining
\begin{equation}
\phi:=\Phi^{D-2} \quad,\quad g_{\mu\nu}:=\phi^{\frac{D-3}{D-2}}\mathbf{g}_{\mu\nu}\quad,\quad V(\phi):=-(D-2)(D-3)\phi^{-1/D-2}   \label{ConformalDefsHD}
\end{equation}
and introducing Lagrange multipliers $X_a$ as before to implement
torsion-freedom, this action is seen to be of the form
(\ref{FOGaction}):
\[
S=-\frac{\mathcal{O}_{D-2}}{8\pi G}\int_M X_ade^a+X_a\varepsilon^a\,_b\omega\wedge e^b
+\phi d\omega +\frac{1}{2}V(\phi)\varepsilon  \,.
\]
Further, integrating by parts and discarding boundary terms we have
\[
S=-\frac{\mathcal{O}_{D-2}}{8\pi G}\int_M e^a\wedge dX_a +\omega\wedge d\phi
+X_a\varepsilon^a \,_b\omega\wedge e^b +\frac{1}{2}V(\phi)\varepsilon\,.
\]

If we collect the zero- and one-forms appearing in the last equation into the multiplets
\begin{align}
(\mathbf{X}^i)&:=(X^a,\phi)=(X^-,X^+,\phi) \notag \\
(\mathbf{A}_i)&:=(e_a,\omega)=(e^+,e^-,\omega)  \notag
\end{align}
and use the Poisson bivector
\[
   \mathcal{P}^{ij}=\begin{pmatrix}
                            0     &  -V/2   &  -X^-  \\
	   	            V/2   &    0    &   X^+  \\  
		            X^-   &   -X^+  &   0
    	            \end{pmatrix}\,,
\]	
the action finally takes the form of a Poisson sigma model:
\begin{equation} \label{SSPSM}
S=-\frac{\mathcal{O}_{D-2}}{8\pi G}\int_M \mathbf{A}_i\wedge d\mathbf{X}^i+\frac{1}{2}\mathcal{P}^{ij}\mathbf{A}_i \wedge \mathbf{A}_j
\end{equation}
with a three-dimensional target space, and $i\in \{-,+,3\}$.

From here on we specialize to four dimensions for which
\begin{equation}
\phi:=\Phi^2 \quad,\quad
g_{\mu\nu}:=\sqrt{\phi}\mathbf{g}_{\mu\nu}\quad,\quad
V(\phi):=-\frac{2}{\sqrt{\phi}} \label{ConformalDefs}
\end{equation}

%%%%%%%%%%%%%%%%%%%%%%%%%%%%%%%%%%%%%%
%%%% CANONICAL FORM OF PSM %%%%%%%%%%%
%%%%%%%%%%%%%%%%%%%%%%%%%%%%%%%%%%%%%%

\subsubsection{Canonical Form}   \label{secPSMcanonical}
The action for a Poisson sigma model is already in first order
form. Using coordinates $(t,x)$ on $M$, (\ref{SSPSM}), specialized to four dimensions, reads:
\begin{equation} \label{canonicalPSM}
S=\frac{1}{2G}\int dt \int dx[A_i\dot{\mathbf{X}}^i-\Lambda_i(\mathbf{X}^i\,'+\mathcal{P}^{ij}A_j)]
\end{equation}
with
\begin{equation}
  A_i:=\mathbf{A}_{xi} \quad,\quad
  \Lambda_i:=\mathbf{A}_{ti} \quad,\quad
  \dot{\mathbf{X}}:=\partial_t\mathbf{X} \quad , \quad \mathbf{X}':=\partial_x\mathbf{X}\,.
\end{equation}
The canonically conjugate variables are thus $\mathbf{X}^i$ and $A_i$ with
\[
\{\mathbf{X}^i(x),A_i(y)\}=2G\delta^i_j\delta(x-y)\,,
\]
subject to the total constraint
\[
\int dx \, \Lambda_i \tG^i\approx0
\]
with Lagrange multipliers $\Lambda_i$ and
\[
\tG^i:=\frac{1}{2G}(\mathbf{X}^i\,'+\mathcal{P}^{ij}A_j)\,.
\]
These constraints form the first class algebra
%\footnote{Note the double use of $G$ to denote Newton's gravitational constant and the total PSM constraint, its meaning here and later on should be clear from the context} 
\begin{equation}
 \{\G[\Lambda_i],\G[K_j]\}= -\frac{1}{2G}\G[\Lambda_i K_l\partial_k {\cal P}^{il}]\,.
\end{equation}
For a linear Poisson tensor, this is an algebra with structure
constants, equivalent to the Gauss constraint of a gauge theory with
structure constants $\partial_k{\cal P}^{ij}$. For non-linear Poisson
tensors, on the other hand, the system has structure functions. The
Lie algebroid formulation of Poisson sigma models, alluded to above,
provides an interesting perspective on systems with structure
functions whose constraints generate the symmetries of a Lie algebroid
rather than a local Lie algebra. Similar interpretations exist for a
large class of algebroid Yang-Mills theories \cite{AYM} or Dirac sigma
models \cite{DSM}.

The action (\ref{FOGaction}) is
space-time diffeomorphism and SO(1,1)-gauge invariant. In particular,
$\tG^3$ is the canonical generator of local gauge
transformations, and the spatial diffeomorphism constraint is the
combination
\[
\tilde{D}:=A_i\tG^i=A_i\mathbf{X}^i\,' \,.
\]

For the relation to variables underlying the loop formulation, it will
be convenient to introduce the SO(1,1) invariant quantities
\begin{equation}
X^2:=X^{-}X^{+} \quad,\quad
e^2:=e_x^-e_x^+ 
\end{equation}
as well as gauge angles $\alpha$ and $\beta$ by
\[
          X^{\pm}=X \exp (\pm\beta) \quad,\quad
          e_x^{\pm}=e \exp (\pm\alpha)\,.
\]
The angles are well-defined as long as $X\not=0$ and $e\not=0$, which
is the case except at horizons. In what follows, we analyze the local
constraint algebra such that global problems of this transformation of
variables do not play a role.  (As expected for an Abelian gauge
transformation, $\G^3[\Lambda]$ then generates
$\alpha\to\alpha-\Lambda$, $\beta\to\beta-\Lambda$,
$\omega_x\to\omega_x+\Lambda'$. In the regions where our change of
variables is valid one can even Abelianize the full constraint system;
see \cite{Strobl}.)

The symplectic structure $\Omega$ in the new variables becomes
\begin{align}
\Omega&=\frac{1}{2G}\int dx \,\dd A_i \wedge \dd \mathbf{X}^i  \notag \\
      &=\frac{1}{2G}\int dx \,\dd(e \exp\alpha)\wedge \dd(X\exp -\beta)+\dd(e \exp -\alpha)\wedge \dd(X\exp \beta)+\dd\omega_x\wedge \dd\phi \notag \\
      &=\frac{1}{2G}\int dx \,\dd e\wedge \dd(2X\cosh(\alpha-\beta))+\dd\alpha\wedge \dd(2Xe\sinh (\alpha-\beta))+\dd\omega_x\wedge \dd\phi \notag \\
      &=\frac{1}{2G}\int dx \,\dd e\wedge \dd Q^e+\dd\alpha\wedge \dd Q^\alpha+\dd\omega_x\wedge \dd\phi \notag
\end{align}
with 
\begin{equation}
Q^e:=2X\cosh(\alpha-\beta) \quad,\quad
Q^\alpha:=2Xe\sinh (\alpha-\beta) \,,
\end{equation}
which provides the new canonically conjugate pairs
\begin{equation}
\{Q^e(x),e(y)\}=\{Q^\alpha(x),\alpha(y)\}=\{\phi(x),\omega_x(y)\}=2G\delta(x,y)\,. \label{PBrackets1}
\end{equation}

We can invert this transformation to find
\begin{equation}
X^\mp=\frac{e Q^e \pm Q^\alpha}{2e}\exp(\mp\alpha)    \label{Xpm0}
\end{equation}
and insert it into the PSM constraints:
\begin{gather}
\tG^\mp=\frac{1}{2G}\left[\left(\frac{e Q^e\pm Q^\alpha}{2e}\right)'
\mp\left(\frac{e Q^e\pm Q^\alpha}{2e}\right)(\omega_x+\alpha')\mp
\frac{1}{2}V(\phi)e\right]\exp (\mp\alpha)   \label{Gmp} \\
\tG^3=\frac{1}{2G}(\phi'+Q^\alpha)\,.   \label{G3}
\end{gather}
By the same combination as before, the diffeomorphism constraint is
\begin{equation}
\tilde{D}=\frac{1}{2G}(eQ^e\,'-Q^\alpha\alpha'+\omega_x\phi')  \label{Dconstraint}\,.
\end{equation}

%%%%%%%%%%%%%%%%%%%%%%%%%%%%%%%%%%%%%%%%%%%%%%%%%%
%%%%%%% SS LOOP VARIABLES SUMMARY %%%%%%%%%%%%%%%%
%%%%%%%%%%%%%%%%%%%%%%%%%%%%%%%%%%%%%%%%%%%%%%%%%%

\subsection{Spherically Symmetric Ashtekar Variables}

Models of loop quantum gravity are also formulated in terms of
vielbein and connection components, but in different ways. The
symplectic coordinate chart in this case consists of the
$\mathfrak{su}(2)$-valued Ashtekar connection $A_a^i(\bar{x})$ and the
densitized triad $E^a_i(\bar{x})$ (which can be seen as an
$\mathfrak{su}(2)$-valued vector field) on a 3d spatial manifold
coordinatized by $\bar{x}$. Their relation to the standard canonical
ADM variables is determined as follows: The spatial 3d metric $q_{ab}$
is constructed from the densitized triad via $(\det
q)q^{ab}=E_i^aE_i^b$, and the Ashtekar--Barbero connection
\cite{AshVar,AshVarReell} is related to the extrinsic curvature
$K^i_a:=(\det E)^{-\frac{1}{2}}K_{ab}E^{bi}$ and the spin connection
$\Gamma^i_a$ compatible with the triad by the formula
\begin{equation} \label{AshtekarConnection}
A^i_a=\Gamma^i_a+\gamma K^i_a\,.
\end{equation}
The real constant $\gamma>0$, called the Barbero-Immirzi parameter
\cite{AshVarReell,Immirzi}, does not play a role classically because
it can be changed by canonical transformations. It will become
important after quantization where transformations which could change
$\gamma$ are not represented unitarily.

The midisuperspace formulation in terms of these variables was
initiated in \cite{SphKl1,SymmRed,SphSymm,SphSymmHam}.  There are
different spherical symmetry types on SU(2)-principal fiber bundles,
only one of which provides non-degenerate triads in its associated
vector bundle \cite{SymmRed}. Using spatial coordinates
$(x,\vartheta,\varphi)$ gives the reduced densitized triad
\[
E=E^x(x)\tau_3\sin\vartheta\partial_x+(E^1(x)\tau_1+E^2(x)\tau_2)\sin\vartheta\partial_\vartheta+(E^1(x)\tau_2-E^2(x)\tau_1)\partial_\varphi
\]
in this non-degenerate sector, which contains only three functions
$E^x$, $E^1$ and $E^2$. Here, we write $E$ as a Lie-algebra valued
field using $\tau_i$ as a basis for $\mathfrak{su}(2)$. The form of
connections $A$ preserving the corresponding symmetry up to gauge is
\[
A=A_x(x)\tau_3dx+(a_1(x)\tau_1+a_2(x)\tau_2)d\vartheta+(a_1(x)\tau_2-a_2(x)\tau_1)\sin\vartheta d\varphi+\tau_3\cos \vartheta d\varphi
\]
and similarly the extrinsic curvature $K$ takes the form
\[
K=K_x(x)\tau_3dx+(K_1(x)\tau_1+K_2(x)\tau_2)d\vartheta+(K_1(x)\tau_2-K_2(x)\tau_1)\sin\vartheta d\varphi
\]

The SU(2)-gauge freedom of the original variables leaves a residual
U(1)-gauge in the reduced theory. As before, we introduce quantities invariant under this Abelian gauge group: $A_x,E^x,K_x$ as they already appear in the fields together with
\begin{equation}
A_\varphi:=\sqrt{a_1^2+a_2^2} \quad,\quad
E^\varphi:=\sqrt{(E^1)^2+(E^2)^2} \quad,\quad
K_\varphi:=\sqrt{K_1\,^2+K_2\,^2}\,.
\end{equation}
The remaining freedom of the four functions $E^1$, $E^2$ and $a_1$,
$a_2$ not contained in $E^\varphi$ and $A_\varphi$ is pure gauge and
can be parameterized by gauge angles $\eta(x)$ and $\tilde{\beta}(x)$,
defined such that
\[
\tau_1\cos \eta+\tau_2\sin \eta:=(E^1\tau_2-E^2\tau_1)/E^\varphi\quad,\quad
\tau_1\cos \tilde{\beta}+\tau_2\sin \tilde{\beta}:=(a_1\tau_2-a_2\tau_1)/A_\varphi\,.
\]
Each of these can be changed by the same amount with a gauge
transformation, so only the difference
$\tilde{\alpha}:=\eta-\tilde{\beta}$ is gauge invariant.

Relation (\ref{AshtekarConnection}) then gives \cite{SphSymmHam}
\begin{equation}
A_\varphi\cos\tilde{\alpha}=\gamma K_\varphi\quad,\quad
A_x+\eta'=\gamma K_x \quad\mbox{and}\quad
A_\varphi^2=\Gamma_\varphi^2+\gamma^2 K_\varphi^2
\end{equation}
with
\begin{equation}
\Gamma_\varphi=-\frac{E^x\,'}{2E^\varphi}\,.  \label{varphiSpinConn}
\end{equation}

The symplectic structure of the original variables
\begin{equation}\label{LoopSymp}
\{A^i_a(\bar{x}),E^b_j(\bar{y})\}=8\pi G\gamma\delta_a^b\delta^i_j\delta(\bar{x},\bar{y})
\end{equation}
makes the functions $E^x$, $E^1$ and $E^2$ canonically conjugate to
$A_x$, $a_1$ and $a_2$, respectively. However, $E^\varphi$ is not
canonically conjugate to $A_\varphi$ due to the non-linear
transformation from the original field components. It turns out that
$E^{\varphi}$ is instead canonically conjugate to $K_{\varphi}$
\cite{SphSymmHam}.  The most suitable polarization for a loop
quatization of spherically reduced gravity turns out to be given
by the canonical pairs:
\begin{equation}\label{SphSymmSymp}
\{A_x(x),\frac{1}{2\gamma}E^x(y)\}=\{K_\varphi(x),E^\varphi(y)\}=\{\eta(x),\frac{1}{2\gamma}P^\eta(y)\}=G\delta(x,y)\,.
\end{equation}

The reduced 2d action is then
\begin{equation} \label{LOOPaction}
S=\int dt \left[\frac{1}{2G\gamma}\int dx \, (E^x \dot{A}_x +
  2\gamma E^\varphi 
\dot{K}_\varphi + P^\eta \dot{\eta})-\int
  dx\,(\lambda\tilde{\mathcal{G}}_{\rm grav} + 
N^x\tilde{\mathcal{D}}_{\rm grav} + N\tilde{\mathcal{H}}_{\rm grav})\right]\,.
\end{equation}
with the Gauss constraint
\begin{equation}
G_{\rm grav}[\lambda]=\frac{1}{2G\gamma}\int dx\,\lambda(E^x\,'+P^\eta)  \label{Gauss}
\end{equation}
generating U(1)-gauge transformations, the diffeomorphism constraint
\begin{equation}
D_{\rm grav}[N^x]=\frac{1}{2G}\int dx\,N^x(2E^\varphi K_\varphi'-K_xE^x\,'+\frac{1}{\gamma}\eta'(E^x\,'+P^\eta))  \label{diffeomorphism}
\end{equation}
generating diffeomorphisms on the one dimensional radial manifold
and the Hamiltonian constraint
\begin{equation} \label{SphSymmHam}
H_{\rm grav}[N]=-\frac{1}{2G}\int dx\,N|E^x|^{-\frac{1}{2}}(K_\varphi^2E^\varphi+2K_\varphi K_xE^x + (1-\Gamma_\varphi^2)E^\varphi+2\Gamma_\varphi'E^x)
\end{equation}
generating dynamical evolution.

%%%%%%%%%%%%%%%%%%%%%%%%%%%%%%%%%%%%%%%%%%%%%%%%%%
%%%%%%%  COMPARING THE TWO MODELS %%%%%%%%%%%%%%%%
%%%%%%%%%%%%%%%%%%%%%%%%%%%%%%%%%%%%%%%%%%%%%%%%%%

\section{Relating the two models}
\label{s:Relation}

The actions (\ref{canonicalPSM}) and (\ref{LOOPaction}) represent
equivalent canonical formulations of spherically reduced general
relativity, so there must exist a canonical transformation between the
PSM and Ashtekar variables. To find such a transformation we first compare
the form of the reduced 2-dimensional metric in terms of these two
sets of variables. This relates the dilaton field $\phi$ and the gauge
invariant part $e$ of the dyad directly to the densitized triad
components $E^x$ and $E^\varphi$. Using this and imposing the
canonical relations (\ref{PBrackets1}) gives a system of differential
equations for the remaining PSM variables. We then use equations of
motion (\ref{spinConn}) to fix some of the ambiguities and check for
consistency.

\subsection{Comparison of metrics}

The general canonical line element
$ds^2=-N^2dt^2+q_{ab}(dx^a+N^adt)(dx^b+N^bdt)$ adapted to spherical
symmetry with coordinates $(t,x,\vartheta,\varphi)$, lapse function
$N(t,x)$ and shift vector $N^x(t,x)$ is
\begin{equation}\label{CanonMetric}
ds^2=-N(t,x)^2dt^2+q_{xx}(t,x)(dx+N^x(t,x)dt)^2 
+q_{\varphi\varphi}(t,x)d\Omega^2
\end{equation}
where $q_{xx}$ and $q_{\varphi\varphi}$ are components of the spatial
metric $dq^2=q_{xx}(t,x)dx^2+q_{\varphi\varphi}(t,x)d\Omega^2$. In
terms of the densitized triad, we have
\[
q_{xx}=\frac{E^\varphi\,^2}{|E^x|}\quad, \quad q_{\varphi\varphi}=|E^x|\,.
\]

Comparing this form of the metric with (\ref{SSmetric}) directly gives
\begin{equation}
q_{\varphi\varphi}=\Phi^2 \quad,\quad
\mathbf{g}_{\mu\nu}=\begin{pmatrix}
                      -N^2+q_{xx}(N^x)^2  &&   q_{xx}N^x \\
                        q_{xx}N^x         &&   q_{xx}  
                     \end{pmatrix} \,.
\end{equation}
This relates the densitized spherically symmetric triad variables to
the dyads of the conformally transformed metric
$g_{\mu\nu}=\sqrt{\phi}\mathbf{g}_{\mu\nu}$ and the dilaton field
$\phi$:
\begin{eqnarray*}
\phi=\Phi^2&=&|E^x|\\
g=2e^+e^-&=&\sqrt{\phi}\mathbf{g}  \\
          \begin{pmatrix}
           2e^+_te^-_t        && e^+_te^-_x+e^+_xe^-_t \\
           e^+_te^-_x+e^+_xe^-_t   && 2e^+_xe^-_x 
          \end{pmatrix}   
         &=&|E^x|^\frac{1}{2}\begin{pmatrix}
                                          -N^2+\frac{E^\varphi\,^2}{|E^x|}N^x\,^2  &&   \frac{E^\varphi\,^2}{|E^x|}N^x \\
                                          \frac{E^\varphi\,^2}{|E^x|}N^x         &&   \frac{E^\varphi\,^2}{|E^x|} 
                                        \end{pmatrix}\,.
\end{eqnarray*}
From this we obtain
\[
e^2=e^+_xe^-_x=\frac{E^\varphi\,^2}{2|E^x|^\frac{1}{2}}
\]
and
\begin{align}
  e^+_x=p\frac{E^\varphi}{\sqrt{2}|E^x|^\frac{1}{4}}\exp\alpha   \quad&,\quad
  e^-_x=p\frac{E^\varphi}{\sqrt{2}|E^x|^\frac{1}{4}}\exp(-\alpha)   \notag \\
  e^+_t=p\frac{N^xE^\varphi\pm N|E^x|^\frac{1}{2}}{\sqrt{2}|E^x|^\frac{1}{4}}\exp\alpha \quad&,\quad
  e^-_t=p\frac{1}{\sqrt{2}|E^x|^\frac{1}{4}}\left(\frac{-N^2|E^x|+N^x\,^2E^\varphi\,^2}{N^xE^\varphi\pm N|E^x|^\frac{1}{2}}\right)\exp(-\alpha) \label{diads1}
\end{align}
with $p=\pm 1$ distinguishing different solutions.

Using equations (\ref{spinConn}) and (\ref{Xpm}) and the equations of motion 
\begin{align}
\dot{E}^x&=2sNK_\varphi|E^x|^\frac{1}{2}+N^xE^x\,'   \notag \\
\dot{E}^\varphi&=N(K_\varphi E^\varphi+K_xE^x)|E^x|^{-\frac{1}{2}}+(N^xE^\varphi)'  \notag 
\end{align}
for the spherically symmetric loop variables where $s$ is the sign of
$E^x$, we get the dependence of the spin connection $\omega_x$ and
Lagrange multipliers $X^\pm$ in terms of
$(E^x,E^\varphi,K_x,K_\varphi)$:
\begin{equation}
\omega_x=\pm s K_x\pm \frac{E^\varphi}{2|E^x|}K_\varphi-\alpha' \label{spinConn2}
\end{equation}
and
\begin{align}
X^{-}&=p\sqrt{2}|E^x|^\frac{1}{4}\left(- s\frac{E^x\,'}{2E^\varphi}\mp K_\varphi\right)\exp (-\alpha)   \label{Xm}\\
X^+&=p\sqrt{2}|E^x|^\frac{1}{4}\left(s\frac{E^x\,'}{2E^\varphi}\mp K_\varphi\right)\exp \alpha\,.   \label{Xp}
\end{align}

%%%%%%%%%%%%%%%%%%%%%%%%%%%%%%%%%%%%%%%%%%%%%%%%%%
%%%%%%% CANONICAL TRANSFORMATION %%%%%%%%%%%%%%%%%
%%%%%%%%%%%%%%%%%%%%%%%%%%%%%%%%%%%%%%%%%%%%%%%%%%

\subsection{Canonical transformation}

We now look for a canonical transformation between the two sets of variables 
\[
(Q^e,Q^\alpha,\phi;e,\alpha,\omega_x) \quad \leftrightarrows \quad (E^x,E^\varphi,P^\eta;A_x,K_\varphi,\eta)\,.
\]
The Poisson bracket relations (\ref{PBrackets1}) give a system of
partial differential equations for the functional dependence of $Q^e$,
$Q^\alpha$, $\alpha$, and $\omega_x$ on the spherically symmetric loop
variables. (There are 15 nontrivial relations that must be
simultaneously satisfied to ensure consistency: $\{Q^e,e\}=2G,
\{Q^e,\phi\}=\{Q^\alpha,e\}=\{Q^\alpha,\phi\}=0,
\{\phi,\omega_x\}=2G, \{e,\omega_x\}=\{\phi,\alpha\}=\{e,\alpha\}=0,
\{Q^\alpha,\alpha \}=2G,
\{Q^e,\alpha\}=\{Q^e,Q^\alpha\}=\{Q^e,\omega_x\}=\{Q^\alpha,\omega_x\}=\{\alpha,\omega_x
\}=0$. The remaining $\{\phi,e\}=0$ is automatically satisfied
given the functional dependence of $\phi$ and $e$ on $E^x$ and
$E^\varphi$.)

These equations are solved in App.~\ref{a:CanTrans}, providing the
canonical transformation
\begin{align}
     Q^e=p2\sqrt{2}|E^x|^\frac{1}{4}K_\varphi+h[|E^x|^{-\frac{1}{4}}E^\varphi] \quad&,\quad
       e=p\frac{E^\varphi}{\sqrt{2}|E^x|^\frac{1}{4}}                          \notag \\
    \phi=|E^x|                                    \quad&,\quad
\omega_x=-sK_x-\frac{E^\varphi}{2|E^x|}K_\varphi+\frac{1}{k}\eta'      \notag \\
Q^\alpha=\frac{k}{\gamma}P^\eta+\left(\frac{k-s\gamma}{\gamma}\right)E^x\,'             \quad&,\quad
  \alpha=-\frac{1}{k}\eta                                         \label{SOLUTION} 
\end{align}
with inverse transformation
\begin{eqnarray}
E^x=s\phi \quad&,&\quad
E^\varphi=p\sqrt{2}\phi^\frac{1}{4}e           \nonumber \\
K_\varphi=p\frac{(Q^e-h)}{2\sqrt{2}\phi^\frac{1}{4}}\quad&,&\quad
K_x=-s(\omega_x+\alpha'+\frac{e}{4\phi}(Q^e-h))         \nonumber \\
\eta=-k \alpha \quad&,&\quad
P^\eta=\frac{\gamma}{k}Q^\alpha+\left(\frac{\gamma-sk}{k}\right)\phi'\,.
   \label{INVSOLUTION}
\end{eqnarray}
Here again, $s={\rm sign}(E^x)$, $k$ is an arbitrary constant, and $h$ an
arbitrary function of one variable.

%%%%%%%%%%%%%%%%%%%%%%%%%%%%%%%%%%%%%%%%%
%%%%%%%%  CONSTRAINTS  %%%%%%%%%%%%%%%%%%
%%%%%%%%%%%%%%%%%%%%%%%%%%%%%%%%%%%%%%%%%
\subsection{Constraints}
We take $h=0$ to provide a specific canonical transformation. As
mentioned in App.~\ref{a:CanTrans}, with this solution the $\G^3$
constraint (\ref{G3}) reproduces the Gauss constraint
(\ref{Gauss}):\footnote{There is a local agreement of the
infinitesimal Gauss symmetries generated by the constraints. Globally,
however, the formulations differ, one having a compact group U(1), the
other the noncompact SO(1,1). In fact, different time gauges have been
used to reduce space-time metrics to objects in canonical form, which
turns out to imply different topological properties of the gauge
orbits.}
\begin{equation}
\G^3[\lambda]=\frac{1}{2G}\int dx\, \lambda(\phi'+Q^\alpha)
            =\frac{k}{2G\gamma}\int dx\,\lambda(E^x\,'+P^\eta) 
           =k\,G_{\rm grav}[\lambda]\,.
\end{equation}
The diffeomorphism constraint (\ref{Dconstraint})  reads: 
\begin{align}
D[N^x]&=\frac{1}{2G}\int dx\, N^x(e Q^e\,'-Q^\alpha \alpha'+\omega_x\phi') \notag \\
      &=\frac{1}{2G}\int dx\,N^x(2E^\varphi K_\varphi'-K_xE^x\,'+
\gamma^{-1}\eta'(P^\eta+E^x\,')) =D_{\rm grav}[N^x]\,.
\end{align}
Using (\ref{varphiSpinConn}), the remaining independent linear
combination becomes
\begin{align}
\G^+[\frac{\sqrt{2}}{2}N\phi^{1/4} \exp(-\alpha)]-\G^-[\frac{\sqrt{2}}{2}N\phi^{1/4} \exp(\alpha)]&=
\frac{\sqrt{2}}{4G}\int dx\,N\phi^{1/4}[Q^e(\omega_x+\alpha')-\left(\frac{Q^\alpha}{e}\right)'
+Ve]\notag\\
=\frac{p}{2G}\int dx\,N 
%|E^x|^{-\frac{1}{4}}
\bigg[&-|E^x|^{-\frac{1}{2}}K_\varphi^2 E^\varphi - 2s |E^x|^{\frac{1}{2}}K_x K_\varphi + \frac{E^\varphi V}{2} \notag \\ 
                     &+\frac{|E^x|^{-\frac{1}{2}} E^x\,'^2}{4E^\varphi}-\frac{s |E^x|^{\frac{1}{2}}E^x\,' E^\varphi\,'}{E^\varphi\,^2}+\frac{s |E^x|^{\frac{1}{2}} E^x\,''}{E^\varphi} \bigg]  \notag \\
-\frac{pk}{2G\gamma}\int
dx\,N|E^x|^{1/4}\bigg[&\frac{|E^x|^{\frac{1}{4}}}{E^\varphi}(E^x\,'+
  P^\eta)\bigg]' \label{DilatonHam}
\end{align}
and reproduces the Hamiltonian constraint (\ref{SphSymmHam}) with
$V(\phi)=-\frac{2}{\sqrt{\phi}}=-2|E^x|^{-\frac{1}{2}}$ (up to the
Gauss constraint):
\[
\G^+[\frac{\sqrt{2}}{2}N \phi^{1/4}\exp(-\alpha)]-
\G^-[\frac{\sqrt{2}}{2}N \phi^{1/4}
\exp(\alpha)]=p H_{\rm grav}[N
% |E^x|^{-\frac{1}{4}}
]-
\frac{pk}{2G\gamma}\int dx\,N|E^x|^{1/4}
\bigg[\frac{|E^x|^{\frac{1}{4}}}{E^\varphi}
(E^x\,'+ P^\eta)\bigg]'\,.
\]

To summarize,
\begin{align}
\tG^3&=k\tilde{\mathcal{G}}_{\rm grav}  \label{G3Gg} \\
e\,\exp(\alpha)\tG^-+e\,\exp(-\alpha)\tG^++\omega_x\tG^3&=
\tilde{\mathcal{D}}_{\rm grav} \label{DDg} \\
-\exp(\alpha)\tG^-+\exp(-\alpha)\tG^++\left(\frac{\tG^3}{e}\right)'&=
\sqrt{2}p\phi^{-\frac{1}{4}}\tilde{\mathcal{H}}_{\rm grav} \label{GmpHg}
\end{align}
verifying once more that (\ref{canonicalPSM}) and (\ref{LOOPaction})
represent equivalent constrained systems.

%%%%%%%%%%%%%%%%%%%%%%%%%%%%%%%%%%%%%%%%%%%
%%%%%%%%% YANG-MILLS FIELDS %%%%%%%%%%%%%%%
%%%%%%%%%%%%%%%%%%%%%%%%%%%%%%%%%%%%%%%%%%%

\section{Inclusion of Yang-Mills fields}
\label{s:YM}

Before discussing our main topic, the role of quantum corrections, we
extend the formalism to include 2-dimensional Yang--Mills fields~\cite{Strobl}. This
will provide a non-trivial model in the presence of quantum
corrections.  The general 2-dimensional Yang--Mills action with an
arbitrary coupling $\zeta$, allowed to depend on the dilaton field
$\phi$, reads
\[
S_{\rm YM}=-\frac{1}{4}\int_M\zeta\, {\rm tr}(\mathcal{F}\wedge*\mathcal{F})
\]
where $\mathcal{F}^I=d{\cal A}^I+\frac{1}{2}c^I\,_{JK}{\cal A}^J\wedge
{\cal A}^K$ is the usual curvature 2-form of the connection ${\cal
A}^IT_I$, with $T_I$ a basis for the Lie algebra $\mathfrak{g}$ with
structure constants $c^{I}{}_{JK}$ of the chosen $n$-dimensional
internal gauge group.  (It should be stressed that from a physical
point of view this is a toy model; spherically reduced Yang--Mills theory
contains extra fields and does not coincide with the purely 2d model.)

In first order form this action is
\begin{equation} \label{FOYMaction}
S_{\rm YM}=-\int_M  {\cal E}^I\mathcal{F}_I+
2\zeta(\phi){\cal E}^I{\cal E}_I\varepsilon\,.
\end{equation}
(Assuming ${\rm tr}(T_IT_J)=\frac{1}{2}\delta_{IJ}$, this equivalence is
seen most easily by inserting the field equation
${\cal E}^I=\frac{1}{4\zeta}*{\cal F}^I$ into the original Yang-Mills action.)
Then $S_{\rm grav}+S_{\rm YM}$ reads
\begin{equation}
S_{\rm gravYM}=-\frac{1}{2G}\int_M X^aDe_a +\phi d\omega + 
2G{\cal E}^I\mathcal{F}_I+\left( \frac{1}{2}V(\phi)+
4G\zeta(\phi){\cal E}^I{\cal E}_I\right)\varepsilon
\end{equation}
where indices $a$ run over $+$ and $-$, and $I=1,\dots, n$. The coupling
of Yang--Mills theory to gravity thus changes the dilaton potential in
a way which depends on the value of the dilaton through the coupling
function $\zeta$. Moreover, the target manifold of the PSM has a
higher dimension due to the degrees of freedom of the Yang--Mills
field: After integrating by parts and dropping the corresponding
surface terms, and with the identifications
\[
(\mathbf{X}^i):=(X^a,\phi,{\cal E}^I) \quad,\quad
(\mathbf{A}_i):=(e_a,\omega,{\cal A}_I)\,,
\]
the previous action turns out to be of Poisson sigma form
(\ref{PSMaction}) on an $(n+3)$-dimensional target space $N$ with
Poisson brackets
\begin{eqnarray*}
\{X^+,X^-\}=V/2+4G\zeta {\cal E}^I{\cal E}_I\quad &,& 
\quad \{X^\pm,\phi\}=\pm X^\pm\\
\{{\cal E}^I,{\cal E}^J\}=c^{IJ}\,_K{\cal E}^K\quad&,&\quad
\{X^\pm,{\cal E}^I\}=\{\phi,{\cal E}^I\}=0\,.
\end{eqnarray*}
The Poisson bivector can thus be decomposed as
\[
(\mathcal{P}^{ij})=\begin{bmatrix}
                    \mathbf{P} & 0 \\
		        0      & \mathbf{Q}
		   \end{bmatrix}
\]
with the $3\times 3$ matrix
\begin{equation} \label{PSMYM}
\mathbf{P}=\begin{pmatrix}
            0                           &   -V/2-4G\zeta {\cal E}^I{\cal E}_I   &  -X^-  \\
	    V/2+4G\zeta {\cal E}^I{\cal E}_I           &    0                   &   X^+  \\  
	    X^-                         &   -X^+                 &   0
	   \end{pmatrix}		   	         
\end{equation}
and the $n\times n$ matrix
\[
\mathbf{Q}=(Q^{IJ})=(c^{IJ}\,_K{\cal E}^K)\,.
\]

The canonical formulation of Sec.~\ref{secPSMcanonical} proceeds
almost unchanged: The symplectic structure (\ref{LoopSymp}) is
extended by the Yang-Mills pairs
\[
\{{\cal E}^I(x),{\cal A}_{I}(y)\}=\delta(x,y)\,.
\]
The constraints $\G^\mp$ (\ref{Gmp}) (and (\ref{GmpT})) receive additional
terms $\mp 2\zeta {\cal E}^I{\cal E}_I e\exp(\mp\alpha)$:
\begin{equation} \label{GmpYM}
\tG^\mp=\frac{1}{2G}\left[\left(\frac{e Q^e\pm Q^\alpha}{2e}\right)'
\mp\left(\frac{e Q^e\pm Q^\alpha}{2e}\right)(\omega_x+\alpha')\mp
\frac{1}{2}V(\phi)e \mp 4G\zeta {\cal E}^I{\cal E}_Ie\right]\exp (\mp\alpha) 
\end{equation}
and we have the usual $n$-component Gauss constraints
\[
\tG^{3+I}={\cal E}^I\,'+c^{IJ}_K {\cal A}_J {\cal E}^K
\]
for the Yang--Mills part of the theory.

\section{Loop quantization of general 2-dimensional dilaton gravity}
\label{s:Loop}

We can now obtain the first application of our canonical relation
between dilaton gravity and spherically symmetric gravity in Ashtekar
variables: a loop quantization of general 2-dimensional dilaton
gravity models. So far in this context, loop quantizations have only
been performed for one specific dilaton potential $V(\phi)\propto
1/\sqrt{\phi}$, corresponding to spherically symmetric gravity in four
dimensions \cite{SphSymm}. Looking at the Hamiltonian constraint in
the form (\ref{DilatonHam}) shows that the potential appears only at
one place, which is in fact a rather simple term in the Hamiltonian
(resulting from the term $E^{\varphi}$ in the parenthesis in
(\ref{SphSymmHam}), which is the only term independent of
$K_{\varphi}$ or $\Gamma_{\varphi}$). This observation allows us
immediately to extend the existing loop quantization of spherically
symmetric gravity to an arbitrary potential $V(\phi)$, provided only
that the expression $V(\phi)$ can be turned into a well-defined
operator. Since this is merely a function of the triad component
$E^x$, such quantizations easily exist.

\subsection{Loop quantization}

To quantize spherically symmetric gravity, we have to define a
well-defined algebra of quantities which seperate points on the
classical phase space of the fields $(A_x,K_{\varphi},\eta;
E^x,E^{\varphi},P^{\eta})$. In particular, some of the fields must be
integrated (``smeared'') in suitable ways so as to provide an algebra
under taking Poisson brackets free of the delta functions which appear
for the fields in (\ref{SphSymmSymp}). The resulting algebra is then
well-defined and can be represented on a Hilbert space to provide the
basic representation of the quantum theory. In gravitational systems,
the smearing must be done with care because there is no background
metric to define the integrations (in addition to the physical metric
given by $E^x$ and $E^{\varphi}$ which are to be quantized).

A loop quantization is based on holonomies
\begin{equation}
h_e[A_x] = \exp\left(\tfrac{1}{2}i \smallint_e A_xd x\right)
\quad,\quad
 h_v[K_{\varphi}] =\exp(i \gamma K_{\varphi}(v))\quad,\quad
h_v[\eta]=\exp(i\eta(v))
\end{equation}
as smeared versions of the configuration variables.  Here, we have
used arbitrary curves $e$ and points $v$ in the radial line as
labels. Varying $e$ and $v$ allows one to recover smooth local fields
unambiguously. The appearance of integrations without auxiliary
structures is dictated by the tensorial nature of the variables: The
U(1)-connection $A_x$ can naturally be integrated to define parallel
transport, while the remaining components are scalars which we simply
exponentiate without integrations. In this framework, exponentiations
are not strictly necessary, but we use them in order to take into
account the origin of these objects from non-Abelian holonomies in the
full setting. Using exponentials instead of linear expressions in
connection or extrinsic curvature components will not spoil the linear
nature of the underlying basic algebra; see Eq.~(\ref{HolFluxAlg}) below.

Similarly, we define flux variables
\begin{equation}\label{hx}
F_v[E^x] = E^x(v)
\quad,\quad
 F_e[E^{\varphi}] = \int_e E^{\varphi}d x \quad,\quad
F_e[P^{\eta}]=\int_e P^{\eta}d x
\end{equation}
for the momenta. Also here, the integrations are naturally dictated by
transformation properties of the fields, $E^x$ being scalar while
$E^{\varphi}$ and $P^{\eta}$ are densities of weight one. Without
introducing a background metric or integration measure, we have thus
managed to integrate all fields such that a well-defined algebra
results:
\begin{eqnarray}
 \{h_e[A_x],F_v[E^x]\}= i\gamma G \delta_{v\in e}
 h_e[A_x]\nonumber\\
\{h_v[K_{\varphi}], F_e[E^{\varphi}]\}= i\gamma G \delta_{v\in e}
 h_v[K_x]\label{HolFluxAlg}\\
\{h_v[\eta], F_e[P^{\eta}]\}= 2i\gamma G \delta_{v\in e}
 h_v[\eta]\nonumber
\end{eqnarray}
where $\delta_{v\in e}$ is one if $v\in e$ and zero otherwise.

An irreducible representation of this algebra can easily be
constructed.  In the connection representation (which is customarily
used in the full theory), an orthonormal basis of states is given by
\begin{equation}
 T_{g,k,\mu}[A_x,K_{\varphi},\eta]=\prod_{e\in g} \exp\left(\tfrac{1}{2}i k_e
\smallint_eA_x d x\right)  \prod_{v\in g}
\exp(i\mu_v \gamma K_{\varphi}(v)) \exp(ik_v\eta(v))
\end{equation}
with integer labels $k_e$, $k_v$ and positive real labels $\mu_v$ on
edges $e$ and vertices $v$, respectively, forming a finite graph $g$
in the 1-dimensional radial line. The labels determine the connection
dependence by irreducible representations of the groups spanned by the
holonomies. (These groups are U(1) for $A_x$- and $\eta$-holonomies
and the Bohr compactification $\bar{\mathbb R}_{\rm Bohr}$ of the real
line for $K_{\varphi}$-holonomies; see \cite{SphSymm} for details.)

Holonomies then simply act as multiplication operators. Specifically:
\begin{equation}
 \hat{h}_e[A_x]
 T_{g,k,\mu}= T_{g,k+\delta_e,\mu}\quad,\quad
 \hat{h}_v[K_{\varphi}] T_{g,k,\mu}= T_{g,k,\mu+\delta_v}
 \quad,\quad \hat{h}_v[\eta] T_{g,k,\mu}= T_{g,k+\delta_v,\mu}
\end{equation}
where $\delta_e$ is a function on the set of all edges which is one
for the edge $e$ and zero otherwise, and analogously for the function
$\delta_v$ on the set of vertices. The state
$T_{g,0,0}[A_x,K_{\varphi},\eta]=1$ is cyclic for this representation.

The densitized triad components, which are momenta conjugate to the
connection components, act as derivative operators:
\begin{eqnarray}
 \hat{F}_v[E^x] T_{g,k,\mu} &=& \gamma\ell_{\rm P}^2
\frac{k_{e^+(v)}+k_{e^-(v)}}{2} T_{g,k,\mu} \label{Exspec}\\
\hat{F}_e[E^{\varphi}]T_{g,k,\mu} &=& \gamma\ell_{\rm P}^2
\sum_{v\in e} \mu_v T_{g,k,\mu}\label{Epspec}\\ 
\hat{F}_e[P^{\eta}] T_{g,k,\mu}&= & 2\gamma\ell_{\rm P}^2 \sum_{v\in e} 
k_v T_{g,k,\mu}
\end{eqnarray}
where $\ell_{\rm P}^2=G\hbar$ is the Planck length squared and
$e^{\pm}(v)$ denote the edges neighboring a point $v$,
distinguished from each other using a given orientation of the radial
line. (We have $k_{e^+(v)}=k_{e^-(v)}$ if $v$ is not a vertex of the
graph.)  All flux operators have discrete spectra: eigenstates as seen
in (\ref{Exspec}) and (\ref{Epspec}) are normalizable. But only
$\hat{E}^x$ has a discrete set of eigenvalues, while
$\hat{E}^{\varphi}$-eigenvalues fill the real line. (Their eigenstates
are elements of the non-separable Hilbert space of square integrable
functions on the Bohr compactification of the real line.)

These basic operators can be used for composite operators as well,
providing well-defined but rather complicated constraint operators.
The Gauss constraint is linear in triad components and can directly be
quantized in terms of the basic flux operators and implies
\begin{equation} \label{kGauss}
 k_v=-\tfrac{1}{2}(k_{e^+(v)}-k_{e^-(v)})\,.
\end{equation}
A basis for gauge invariant states in the kernel of the Gauss
constraint is thus
\begin{equation} \label{GaugeInvSpinNetwork}
 \bar{T}_{g,k,\mu}[A_x+\eta',K_{\varphi}]=
\prod_{e\in g} \exp\left(\tfrac{1}{2}i k_e
\smallint_e(A_x+\eta')d x\right)  \prod_{v\in g}
\exp(i\mu_v \gamma K_{\varphi}(v))
\end{equation}
where the labels $k_v$ are eliminated by imposing
(\ref{kGauss}). Accordingly, states solving the Gauss constraint only
depend on $A_x+\eta'$, not on $A_x$ and $\eta'$ separately. The
diffeomorphism constraint can directly be represented in its finite
version, where its action simply moves labelled graphs in the radial
manifold by a spatial diffeomorphism $\Phi$: $\Phi
T_{g,k,\mu}=T_{\Phi(g,k,\mu)}$ where $\Phi(g,k,\mu)=(\Phi(g),k',\mu')$
is the graph $\Phi(g)$ with labels $k'_{\Phi(e)}=k_e$ and
$\mu'_{\Phi(v)}=\mu_v$.

Operators quantizing the Hamiltonian constraint have been constructed,
but their constraint equations for physical states are more difficult
to solve. Several steps are involved in this particular case: First,
we have to quantize the inverse of $E^x$ as it appears in the
classical constraint. There is no direct operator inverse because
$\hat{E}^x$ has a discrete spectrum containing zero. Nevertheless,
well-defined operators with the correct classical limit exist, which
are based on quantizing, e.g., the right hand side of the classical
identity
\[
  4\pi\gamma G\frac{{\rm sgn}(E^x)E^{\varphi}}{\sqrt{|E^x|}}=
  \{A_x,V\}=2i h_e[A_x]\{h_e[A_x]^{-1},V\}
\]
instead of the left hand
side \cite{QSDI}. This can be done using a quantization of the spatial
volume $V=4\pi \int dx E^{\varphi}\sqrt{|E^x|}$ in terms of fluxes and
turning the Poisson bracket into a commutator divided by
$i\hbar$. Secondly, connection and extrinsic curvature components in
the classical constraint are turned into holonomies and then quantized
directly using the basic operators. Finally, one can quantize the spin
connection terms making use of discretizations of the spatial
derivatives. In this process, a well-defined operator
\begin{eqnarray} \label{HOp}
 \hat{H}[N]\!\! &\!\!\!=\!\!\!&\!\! \frac{i}{2\pi
 G\gamma^3\delta^2\ell_{\rm P}^2} \sum_{v,\sigma=\pm1} \sigma N(v)
 {\rm tr}\biggl(\Bigl(h_{\vartheta}h_{\varphi}h_{\vartheta}^{-1}h_{\varphi}^{-1}
 -h_{\varphi}h_{\vartheta}h_{\varphi}^{-1}h_{\vartheta}^{-1}\\
 &&\qquad\qquad+2\gamma^2\delta^2 (1-\hat{\Gamma}_{\varphi}^2)\tau_3\Bigr)
 h_{x,\sigma}[h_{x,\sigma}^{-1},\hat{V}]\nonumber\\ &&+
 \Bigl(h_{x,\sigma}h_{\vartheta}(v+e^{\sigma}(v))h_{x,\sigma}^{-1}h_{\vartheta}(v)^{-1}
 -h_{\vartheta}(v)h_{x,\sigma}h_{\vartheta}(v+e^{\sigma}(v))^{-1}h_{x,\sigma}^{-1}\Bigr.\nonumber\\
&&\qquad\Bigl.+
 2\gamma^2\delta\smallint_{e^\sigma(v)}\hat{\Gamma}'_{\varphi}\Lambda(v)\Bigr)
 h_{\varphi}[h_{\varphi}^{-1},\hat{V}]\nonumber\\ &&+
 \Bigl(h_{\varphi}(v)h_{x,\sigma}h_{\varphi}(v+e^{\sigma}(v))^{-1}h_{x,\sigma}^{-1}-
 h_{x,\sigma}h_{\varphi}(v+e^{\sigma}(v))h_{x,\sigma}^{-1}h_{\varphi}(v)^{-1}\Bigr.\nonumber\\
&&\qquad\Biggl.\Bigl.+
 2\gamma^2\delta\smallint_{e^{\sigma}(v)}\hat{\Gamma}'_{\varphi}\bar\Lambda(v)\Bigr)
 h_{\vartheta}[h_{\vartheta}^{-1},\hat{V}]\biggr) \nonumber
\end{eqnarray}
with the correct classical limit results. Here, $\delta$ is a
parameter appearing in the exponent of angular holonomies
$h_{\vartheta}=\exp(\delta K_{\varphi}\tau_1)$ and
$h_{\varphi}=\exp(\delta K_{\varphi}\tau_2)$. We use the
SU(2)-form of holonomies for compactness and for an easier comparison
with the full theory; as matrix elements we have our basic holonomies
$h_v[K_{\varphi}]$ from $h_{\vartheta}$ and $h_{\varphi}$, as well as
$h_{e^{\sigma}}(v)[A_x]$ from $h_{x,\sigma}(v):=
\exp(\int_{e^{\sigma}(v)} A_x \tau_3d x)$. Moreover, matrix elements
of $\Lambda(v):=\tau_1\cos\eta(v)+\tau_2\sin\eta(v)$ and
$\bar{\Lambda}(v):=-\tau_1\sin\eta(v)+\tau_2\cos\eta(v)$ act by
multiplication with holonomies $h_v[\eta]$. The parameter $\delta$ in
angular holonomies may be a function of $E^x$ instead of a constant;
this represents the possibility of lattice refinements
\cite{InhomLattice,LTB} in the symmetry orbits which a full
Hamiltonian constraint operator in general implies since it creates
new edges in a graph by acting with the corresponding holonomies. In
the reduced model, there are no edges in the symmetry orbits, and thus
no direct way exists to implement the creation of such new
edges. Instead, an $E^x$-dependence of $\delta$ can be used to make the
edge parameter $\delta$ depend on the area of the orbit. A value
$\delta$ shrinking with the area would imply that more edges are
created because the coordinate length of each new edge used is smaller
at a larger areal size.

The constraint equation $\hat{H}[N]\psi=0$ for all $N$ can be
formulated as a set of coupled difference equations for states labeled
by the triad quantum numbers $k_e$ and $\mu_v$, which have the form
\begin{eqnarray}\label{DiffEq}
 && ~~
 \hat{C}_{{\rm R}+}(k_-,k_+-2)^{\dagger}\psi(\ldots,k_-,k_+-2,\ldots)
 +\hat{C}_{{\rm
R}-}(k_-,k_++2)^{\dagger}\psi(\ldots,k_-,k_++2,\ldots)\nonumber\\ 
 &&+\hat{C}_{{\rm L}+}(k_--2,k_+)^{\dagger}\psi(\ldots,k_--2,k_+,\ldots)
 +\hat{C}_{{\rm L}-}(k_-+2,k_+)^{\dagger}\psi(\ldots,k_-+2,k_+,\ldots)
\nonumber\\ 
 &&+\hat{C}_0(k_-,k_+)^{\dagger}\psi(\ldots,k_-,k_+,\ldots)=0\,,
\end{eqnarray}
one for each vertex. Only the edge labels $k_e$ are written explicitly
in this difference expression, but states also depend on vertex labels
$\mu_v$ on which the coefficient operators $\hat{C}_I$ act; see
\cite{SphSymmHam} for details. 
We will only require the central coefficient
\begin{eqnarray} \label{C0}
\hat{C}_0|\vec{\mu},\vec{k}\rangle &=&
\frac{\ell_{\rm P}}{2\sqrt{2}G\gamma^{3/2}\delta^2}\left(
|\mu|\left(\sqrt{|k_++k_-+1|}-\sqrt{|k_++k_--1|}\right)\right.\\
&&\times\left(|\mu_-,k_-,\mu+2\delta,k_+,\mu_+\rangle+ 
|\mu_-,k_-,\mu-2\delta,k_+,\mu_+\rangle\right.\nonumber\\
 &&\quad-  2(1+2\gamma^2\delta^2(1-\Gamma_{\varphi}^2(\vec{\mu},\vec{k})))
  |\mu_-,k_-,\mu,k_+,\mu_+\rangle)\nonumber\\
 &&\left. -4\gamma^2\delta^2{\rm sgn}_{\delta/2}(\mu)  \sqrt{|k_++k_-|}
 \Gamma_{\varphi}'(\vec{\mu},\vec{k})
  |\mu_-,k_-,\mu,k_+,\mu_+\rangle\right) \nonumber\\
 && +\hat{H}_{{\rm
    matter},v}|\mu_-,k_-,\mu,k_+,\mu_+\rangle \nonumber
\end{eqnarray}
below.  For these basic difference equations, one can show that they
are free of gravitational singularities in the sense of quantum
hyperbolicity \cite{BSCG}: the recurrence scheme they define provides
an evolution of the wave function which does not stop where a
classical singularity would form \cite{SphSymmSing}.

These difference equations are linear but quite involved; moreover,
they are formulated for a wave function which is not easy to interpret
in a generally covariant system. Instead of solving the equations
directly, one can make use of an effective analysis which provides
approximate equations of a simpler form for expectation values and
possibly higher moments of the state. These equations are obtained as
a Hamiltonian system whose Hamiltonian function is the expectation
value of the Hamiltonian operator in a general state \cite{EffAc}, and
which are constrained by expectation values of products of constraint
operators \cite{EffCons}. Through the expectation value,
characteristic features of the quantized Hamiltonian, such as
deviations of the quantized $1/E^x$ from the classical behavior, enter
the effective Hamiltonian and give rise to potential physical effects
(e.g.\ for black hole collapse in \cite{LTB}). We will later use this
type of corrections in an analysis compared to a PSM.

\subsection{Applications}

Ashtekar variables only exist in three and four dimensions, such that
an immediate loop quantization is possible only in those cases or in
models which are obtained from them by symmetry reduction. With the
formulation of general dilaton gravity models in spherically symmetric
Ashtekar variables, however, we are now in a position to extend the
loop quantization to arbitrary dilaton potentials: The Gauss and
diffeomorphism constraint remain unaffected; in the Hamiltonian
constraint, we simply insert the appropriate operator quantizing the
given dilaton potential $V(\phi)$ in the spherically symmetric
constraint operator; this will only change the coefficient
$\hat{C}_0$, (\ref{C0}), in a straightforward way: Instead of
$1-\Gamma_{\varphi}^2$ we then have
$-V(E^x)\sqrt{|E^x|}-\Gamma_{\varphi}^2$, where $E^x$ in a triad
representation will simply be replaced by $\frac{1}{2}\gamma\ell_{\rm
P}^2(k_++k_-)$ after quantization. All the other coefficients in the
difference equation remain unchanged, and so does the conclusion about
the absence of singularities. Thus, all dilaton gravity models are
singularity-free in a loop quantization.

This allows several specific applications. First, we can choose a
linear dilaton potential $V(\phi)\propto\phi$, which provides a loop
quantization of $BF$-theory. As one can see, the quantization does not
simplify considerably in this case because most terms of the
Hamiltonian constraint remain unchanged compared to spherically
symmetric gravity. This is quite unexpected given that $BF$-theory can
be quantized rather easily in different formulations. However,
transformations between Ashtekar-type variables and variables which
allow simple quantizations are non-trivial. Their quantizations can
thus differ considerably. The $BF$-case of PSMs, quantized in Ashtekar
variables as obtained here, can provide an interesting model for a
quantization of a simple classical theory, quantized using techniques
and basic objects as they apply to full gravity.

Secondly, we can provide loop quantizations of spherically symmetric
models in arbitrary $D$ space-time dimensions. Here, we insert
quantizations of the corresponding potentials $V(\phi)\propto
\phi^{d(D)}$ which $d(D)=-1/(D-2)$ from (\ref{ConformalDefsHD}). Even
though loop quantizations of general $D$-dimensional theories with
$D>4$ have not been performed, at least non-rotating black holes can
be studied by these models and compared with results from alternative
quantizations in higher dimensions.

%%%%%%%%%%%%%%%%%%%%%%%%%%%%%%%%%%%%%%%%%%%%%
%%% DEFORMATIONS OF THE CONSTRAIT ALGEBRA %%%
%%%%%%%%%%%%%%%%%%%%%%%%%%%%%%%%%%%%%%%%%%%%%

\section{Inverse triad quantization as a consistent deformation}
\label{s:Consistent}

We are now ready to introduce a specific type of quantum corrections
to the classical models considered so far. This type of corrections
results from quantizations of inverse powers of triad components in
the loop formulation, as they appear in the Hamiltonian
constraint. Densitized triads in loop quantum gravity are quantized by
flux operators which have discrete spectra containing zero as an
eigenvalue. Thus, they do not possess densely defined inverse
operators. Nevertheless, as sketched before one can quantize inverse
densitized triad components in a well-defined way \cite{QSDV},
providing densely defined quantum constraint operators. For small
values of the triad components near the classical divergence at zero,
however, expectation values of the inverse triad operators in coherent
states differ from the classical expression of the inverse. This
deviation is captured by introducing a quantum correction function in
terms of the constraint where inverse triad components appear via
properties of triad operators in the loop representation. This type of
correction is directly related to the underlying discreteness of
geometric spectra.

Specifically, we have an inverse $E^x$ multiplying all terms in the
Hamiltonian constraint (\ref{SphSymmHam}) in Ashtekar variables. A
quantum correction function $f$ is thus introduced into the
Hamiltonian constraint as
\begin{equation}
H_{\rm grav}^Q[N]=\int dx\,N f\tilde{\mathcal{H}}_{\rm grav}   \label{QHamiltonian}
\end{equation}
where $H_{\rm grav}[N]=\int dx\,N\tilde{\mathcal{H}}_{\rm grav}$ is
the classical Hamiltonian (\ref{SphSymmHam}). (See \cite{LTB} for
specific examples of correction functions.) The Gauss and
diffeomorphism constraints remain unaltered since they do not contain
inverse triad components.

Assuming that $f$ depends only on the densitized triad components
$E^x$ and $\Ef$ (but not on their spatial derivatives and not on
connection components) the full constraint algebra is
\begin{equation}
\{G_{\rm grav}[\lambda_1],G_{\rm grav}[\lambda_2]\}=0                                         \notag
\end{equation}
\begin{equation}
\{G_{\rm grav}[\lambda],D_{\rm grav}[N^x]\}=-G_{\rm grav}[\mathcal{L}_{N^x}\lambda]=-G_{\rm grav}[N^x\lambda']  \notag 
\end{equation}
\begin{equation}
\{G_{\rm grav}[\lambda],H_{\rm grav}^Q[N]\}=0                                                 \notag
\end{equation}
\begin{equation}
\{D_{\rm grav}[N^x],D_{\rm grav}[M^x]\}=D_{\rm grav}[[\bar{N},\bar{M}]]=D_{\rm grav}[N^xM^x\,'-M^xN^x\,']           \notag 
\end{equation}
\begin{equation}
\{H_{\rm grav}^Q[N],D_{\rm grav}[N^x]\}=-H_{\rm grav}^Q[N^xN'-\frac{1}{f}\frac{\partial f}{\partial\Ef}\Ef N N^x\,'] \notag 
\end{equation}
\begin{equation}
\{H_{\rm grav}^Q[N],H_{\rm grav}^Q[M]\}=D_{\rm grav}[f^2\frac{E^x}{\Ef\,^2}(NM'-MN')]+G_{\rm grav}[f^2\frac{E^x}{\Ef\,^2}(NM'-MN')\eta']\,.  \label{QPalgebra1}
\end{equation}
Thus, any correction of this form provides a first-class algebra, even
though coefficients in the algebra are corrected compared to the
classical case. We thus have a consistent deformation of the classical
theory where the number of all gauge symmetries is preserved (even
though the algebra does change).  Note that $H_{\rm grav}^Q$
transforms as a scalar only if $f$ is independent of $E^{\varphi}$
since $\Ef$ is the only quantity of density weight one. However, the
vacuum algebra is first class even if $f$ does depend on
$E^{\varphi}$. So far, this result is not surprising because the
correction function $f$ simply multiplies the total Hamiltonian
constraint and could thus be absorbed in the lapse function. (This by
itself could change observable properties, as also discussed in
\cite{Tensor}, because it would still be the classical lapse function
which enters the space-time metric (\ref{CanonMetric}) while the lapse
function entering the Hamiltonian would be corrected by $f$. However,
as far as consistency of the deformation is concerned, the vacuum case
is rather trivial.)

The situation changes if we add matter terms by coupling a
two-dimensional Yang-Mills system as in Sec.~\ref{s:YM},
\[
S_{\rm YM}=\int dt\int dx\, {\cal A}_{xI}\dot{{\cal E}}^I-
{\cal A}_{tI}\tilde{\mathcal{G}}_{\rm YM}-N\tilde{\mathcal{H}}_{\rm YM}
\]
with Gauss constraint
\[
G_{\rm YM}[\Lambda_I]=\int dx\, \Lambda_I( {\cal E}^I\,'+c^I_{JK}{\cal
A}_J{\cal E}_K)
\]
and Hamiltonian
\[
H_{\rm YM}[N]=\int dx\,N\zeta\Ef {\cal E}^I{\cal E}_I\,.
\]

There is no inverse triad component in this Hamiltonian, which
remains uncorrected under the effects studied here. Thus, $f$ can no
longer be absorbed in $N$ for the total constraint. The quantum
corrected gravity-Yang-Mills Hamiltonian is
\[
H^Q[N]=H^Q_{\rm grav}[N]+H_{\rm YM}[N]
\]
resulting in the full constraint algebra
\[
\{G_{\rm YM}[\Lambda_I],G_{\rm YM}[V_J]\}=-G_{\rm YM}[c^{IJK}\Lambda_I V_J]
\]
\[
\{G_{\rm YM}[\Lambda_I],G_{\rm grav}[\lambda]\}=\{G_{\rm YM}[\Lambda_I],D_{\rm grav}[N^x]\}=0
\]
\[
\{H^Q[N],G_{\rm grav}[\lambda]\}=\{H^Q[N],G_{\rm YM}[\Lambda_I]\}=0
\]
\begin{align} 
\{H^Q[N],D_{\rm grav}[N^x]\}=&-H_{\rm grav}^Q[N^xN'-\frac{1}{f}\frac{\partial f}{\partial\Ef}\Ef N N^x\,']  \notag  \\
                    &-H_{\rm YM}[N^xN']-G_{\rm YM}[2NN^x\zeta\Ef E_I] \notag
\end{align}
\begin{align}
\{H^Q[N],H^Q[M]\}=&D_{\rm grav}[f^2\frac{E^x}{\Ef\,^2}(NM'-MN')]+G_{\rm grav}[f^2\frac{E^x}{\Ef\,^2}(NM'-MN')\eta']\,.   \label{QPalgebra2}
\end{align}
In contrast to the vaccum case, there is now a non-trivial condition
for the correction function: Only when $f$ does not depend on
$E^{\varphi}$ can all the terms in $\{H^Q[N],D_{\rm grav}[N^x]\}$ be
combined to constraints. The dependence on $E^x$, on the other hand,
is unrestricted. Thus, quantum corrections due to the loop
quantization can provide non-trivial consistent deformations.

On the other hand, it was proved in \cite{Izawa} that a consistent
deformation of the PSM in the sense of \cite{Barnich} must always be a
PSM with the same dimension.  Since the corrected constraint algebras
(\ref{QPalgebra1}) and (\ref{QPalgebra2}) remain first class, the
number of gauge symmetries does indeed stay fixed. It must be possible
to formulate the quantum corrected system as a PSM. However, the
result that any consistent deformation of a PSM must again be a PSM,
as it follows from a BRST analysis, is obtained for equivalence
classes of theories up to field redefinitions. This does not provide a
constructive procedure to determine a corresponding PSM formulation
for a given consistent deformation, and thus further input is
required.

Rewriting the PSM constraints in terms of the
standard gravitational constraints by inverting (\ref{G3Gg}),
(\ref{DDg}) and (\ref{GmpHg}):
\begin{equation} \label{GmpFromGravity}
\tG^\mp=\left[\frac{1}{2e}\tilde{\mathcal{D}}_{\rm grav}\mp 
\frac{p}{\sqrt{2}}\phi^{-\frac{1}{4}}\tilde{\mathcal{H}} - 
\frac{\omega_x}{2e}\tG^3\pm\left(\frac{\tG^3}{2e}\right)'\right]\exp(\mp\alpha)
\end{equation}
(with $\tilde{\mathcal{H}}$ either $\tilde{\mathcal{H}}_{\rm grav}$ or
$\tilde{\mathcal{H}}_{\rm grav}+\tilde{\mathcal{H}}_{\rm YM}$) and
susbstituting $\tilde{\mathcal{H}}_{\rm grav}$ for the quantum
corrected Hamiltonian $f\,\tilde{\mathcal{H}}_{\rm grav}$, we obtain a
deformation of the PSM. It must be possible to cast the anomaly free
algebras (\ref{QPalgebra1}) and (\ref{QPalgebra2}) as a PSM of some
form. Finding this form will provide an action formulation for the
quantum corrected system, and thus a covariant interpretation of the
quantum correction function.

Inserting the correction function $f$ directly in
(\ref{GmpFromGravity}) gives explicitly, in terms of PSM variables,
\begin{equation}
\tG^\mp_{\rm
deformed}=\tG^\mp\mp(f[\phi]-1)\frac{1}{2G}\left[\frac{1}{2}V(\phi)e+
\frac{Q^e}{2}(\omega_x+\alpha')+\left(\frac{\phi'}{2e}\right)'\right]\exp
(\mp\alpha)
\end{equation}
with $\tG^\mp$ as in (\ref{Gmp}) or (\ref{GmpYM}). This is not yet of a
form suitable for a PSM interpretation due to the extra terms
involving e.g.\ derivatives of $\phi$ which cannot simply be put in
the dilaton potential. (The potential must be a function on the target
space, which cannot accomodate space-time derivatives.)

Instead, we can use the requirement of the PSM form to find the
corresponding formulation. In the previous equation, we have simply
taken the same combinations of loop variable constraints as in the
classical case. But if the constraints are corrected, we may well have
to use different combinations of the constraints, with corrected
coefficients, to bring them in a PSM form. We thus change the
coefficients in front of the gravity constraints on the right hand
side of (\ref{GmpFromGravity}) so as to exactly cancel the unwanted
terms depending on $\phi$-derivatives. For this there is a unique way
up to a total factor: the coefficient of $\tilde{\cal H}=f\tilde{\cal
H}_{\rm grav}+ \tilde{\cal H}_{\rm YM}$ in the combination
$\tG^{\mp}$ of constraints must carry an extra factor of
$1/f(\phi)$. In this way, the $\phi$-derivatives cancel in the
combination of constraints as they do classically. The system is then
described by a Poisson sigma model with constraints
\[
\tG^\mp_Q=\left[\frac{1}{2e}\tilde{\mathcal{D}}_{\rm grav}\mp 
\frac{p}{\sqrt{2}}\phi^{-\frac{1}{4}}\left(\tilde{\mathcal{H}}_{\rm grav}+
\frac{1}{f[\phi]}\tilde{\mathcal{H}}_{\rm YM}\right) - \frac{\omega_x}{2e}
\tG^3\pm\left(\frac{\tG^3}{2e}\right)'\right]\exp(\mp\alpha)\,.
\]
Here, the correction function appears only in one place multiplying
the Yang--Mills Hamiltonian. The correction is thus non-trivial and
changes the coupling of Yang--Mills to gravity: We now have the
effective potential
\begin{equation}
 \frac{1}{2}V(\phi)+4G\frac{\zeta(\phi)}{f(\phi)} {\cal E}^I{\cal E}_I\,.
\end{equation}
In these models, the arbitrariness of $\zeta$ (in a $\phi$-dependent
way) is thus enough to account for our consistent deformations: the
deformed Poisson sigma model for (\ref{PSMYM}) is of the same type
with $\zeta$ replaced by $\zeta/f$. This is in accordance with our
condition for a consistent deformation derived from
Eq.~(\ref{QPalgebra2}), namely that $f$ only depends on $E^x$ which is
identified with the dilaton $\phi$. Any other dependence could not be
combined with the Yang--Mills coupling function $\zeta(\phi)$.

%%%%%%%%%%%%%%%%%%%%%%%%%%%%%%%%%%%%%%%%%%%%%%%%%%%%%%%%%%%%%%%%%%%%%%%%%%%%%%%%%%%%%%%%%

\section{Conclusions}

We have studied the canonical relation between 2-dimensional dilaton
gravity, Poisson sigma models and spherically symmetric gravity in
Ashtekar variables. This is of interest because Ashtekar variables
allow a background independent quantization of the full theory, while
other quantization methods have been applied to dilaton gravity in two
dimensions, such as a rigorous path integral quantization. Moreover,
Poisson sigma models allow an interpretation of their structure
functions as defining Lie algebroid symmetries, generalizing the Lie
algebra symmetries of systems with structure constants. Given the
explicit canonical transformation to Ashtekar variables we have
derived, one may ask whether an analogous reformulation as a Lie
algebroid sigma model could exist in four dimensions. If this would be
the case, the structure function issue of general relativity could be
substantially simplified. Unfortunately, the canonical transformation
uses several special features realized only in two dimensions. For
instance, as shown by Eq.~(\ref{DilatonHam}) we need to eliminate a
second derivative of $E^x$ which appears in the Hamiltonian constraint
in Ashtekar variables but not in the PSM constraints. This can be done
in two dimensions by means of a spatial derivative of the Gauss
constraint. In four dimensions, on the other hand, the Gauss
constraint contains the total divergence of the triad, which cannot
provide all terms needed to remove all second triad derivatives from
the full Hamiltonian constraint.

As a side result, we have used some of our derivations to extend the
loop quantization to spherically symmetric systems in arbitrary $D$
space-time dimensions. This extends the proofs of singularity-freedom
of spherically symmetric loop quantizations to spherically symmetric
systems in arbitrary dimensions.

While dilaton gravity in two dimensions has been quantized covariantly
by path integral methods, loop quantum gravity is a canonical
quantization. In this context, the consistency issue of the resulting
quantum constrained system is probably the most important one in loop
quantum gravity, whose analysis will tell whether the diverse effects
studied in simple models can be viable and covariant in general. What
our analysis of consistency in two dimensions has shown is that there
is indeed room for non-trivial effects due to the
quantization. Quantum corrections of the canonical quantization are
then related to a covariant action, where effective couplings to the
Yang--Mills ingredients arose. Similar effects have been studied in
the full four space-time dimensions but with perturbative
inhomogeneities \cite{ConstraintAlgebra}, where consistency was also
shown to be possible.

For further corrections existing in a loop quantization, consistency
has not yet been demonstrated. Among those we especially highlight the
general phenomenon of quantum back-reaction, which implies that
moments of a state such as fluctuations and correlations influence the
dynamical behavior of expectation values. If this is included, new
quantum degrees of freedom arise in an effective theory. In our
context, a consistent deformation of this type will provide a
higher-dimensional target space of the Poisson sigma model. Since the
number of fields changes, the rigidity proofs for consistent
deformations of PSMs no longer apply. Such effective theories could
even generate new algebroid sigma models beyond PSMs, e.g.\ of the
forms introduced in \cite{AYM,DSM}.

\section*{Acknowledgements}

We thank Daniel Grumiller and Thomas Strobl for discussions and
comments. Some of the derivations described here were started while
M.B. visited the Erwin-Schr\"odinger-Institute, Vienna, during the
program ``Poisson Sigma Models, Lie Algebroids, Deformations, and
Higher Analogues,'' whose support is gratefully acknowledged.  This
work was supported in part by NSF grants PHY0653127 and 0748336.

\begin{appendix}

\section{Canonical transformation}
\label{a:CanTrans}

Here, we explicitly compute the canonical relation between Poisson
 sigma models and Ashtekar variables.  The equation $\{Q^e,e\}=2G$
 gives
\[
-\gamma\frac{p s E^\varphi}{4|E^x|^\frac{5}{4}}\frac{\delta Q^e}{\delta A_x}+\frac{p}{2|E^x|^\frac{1}{4}}\frac{\delta Q^e}{\delta K_\varphi}=\sqrt{2}
\]
and $\{Q^e,\phi\}=0$ reads $\delta Q^e/\delta A_x=0$. Together they
imply:
\begin{equation}
Q^e=p2\sqrt{2}\|E^x|^\frac{1}{4}K_\varphi+\tilde{Q}^e[E^x,E^\varphi,P^\eta,\eta]   \label{PBeq1}      %%%%%%%%% PB EQUATION 1 %%%%%%%%%%%%
\end{equation}
for arbitrary function $\tilde{Q}^e[E^x,E^\varphi,P^\eta,\eta]$.

Equations $\{Q^\alpha,e\}=0$ and $\{Q^\alpha,\phi\}=0$ give, respectively,
\[
-\gamma\frac{s E^\varphi}{4|E^x|^\frac{5}{4}}\frac{\delta Q^\alpha}{\delta A_x}+\frac{1}{2|E^x|^\frac{1}{4}}\frac{\delta Q^\alpha}{\delta K_\varphi}=0
\quad,\quad \frac{\delta Q^\alpha}{\delta A_x}=0
\]
which taken together imply
\begin{equation}
Q^\alpha=Q^\alpha[E^x,E^\varphi,P^\eta,\eta]\,.  \label{PBeq2}         %%%%%%%%%%% PB EQUATION 2 %%%%%%%%%%%%%%%%%
\end{equation}
The equation $\{\phi,\omega_x\}=2G$ just implies
\[ 
-\gamma\frac{s}{2}\frac{\delta \omega_x}{\delta A_x}=\frac{1}{2}
\]
and $\{e,\omega_x\}=0$ is equivalent to
\[
\frac{\delta \omega_x}{\delta K_\varphi}=\gamma\frac{s E^\varphi}{2|E^x|}\frac{\delta \omega_x}{\delta A_x}\,.
\]
These two equations give the dependence of $\omega_x$ on $K_x$ and
$K_\varphi$:
\begin{equation}
\omega_x=-\frac{s A_x}{\gamma} - \frac{E^\varphi}{2|E^x|}K_\varphi+\tilde{\omega}_x[E^x,E^\varphi,P^\eta,\eta]  \label{PBeq3}     %%%%%%% PB EQUATION 3 %%%%%%%%
\end{equation}
which fixes the sign ambiguity in (\ref{spinConn2}) and gives
\begin{equation}
\tilde{\omega}_x=-\frac{s \eta'}{\gamma}-\alpha'\,.  \label{PBSpinConntilde}
\end{equation}

Equations $\{\phi,\alpha\}=0$ and $\{e,\alpha\}=0$ are, respectively,
\[
\frac{\delta \alpha}{\delta A_x}=0 \quad,\quad
\frac{\delta \alpha}{\delta K_\varphi}=\gamma\frac{s E^\varphi}{2|E^x|}\frac{\delta \alpha}{\delta A_x}\,.
\]
Thus,
\begin{equation}
\alpha=\alpha[E^x,E^\varphi,P^\eta,\eta]  \label{PBeq4}   %%%%%%% PB EQUATION 4 %%%%%%%%
\end{equation}

The remaining six equations $\{Q^\alpha,\alpha \}=2G$ and
$\{Q^e,\alpha\}=\{Q^e,Q^\alpha\}=\{Q^e,\omega_x\}=\{Q^\alpha,\omega_x\}=\{\alpha,\omega_x
\}=0$ are
\begin{align}
\frac{\delta Q^\alpha}{\delta \eta}\frac{\delta \alpha}{\delta P^\eta}-
\frac{\delta Q^\alpha}{\delta P^\eta}\frac{\delta \alpha}{\delta \eta}&=\frac{1}{\gamma} \label{PBeq5} \\
\frac{\delta \tilde{Q}^e}{\delta \eta}\frac{\delta \alpha}{\delta P^\eta}-
\frac{\delta \tilde{Q}^e}{\delta P^\eta}\frac{\delta \alpha}{\delta \eta}&=-p\frac{\sqrt{2}|E^x|^\frac{1}{4}}{\gamma}\frac{\delta \alpha}{\delta E^\varphi} \label{PBeq6} \\
\frac{\delta \tilde{Q}^e}{\delta \eta}\frac{\delta Q^\alpha}{\delta P^\eta}-
\frac{\delta \tilde{Q}^e}{\delta P^\eta}\frac{\delta Q^\alpha}{\delta \eta}-&=-p\frac{\sqrt{2}|E^x|^\frac{1}{4}}{\gamma}\frac{\delta Q^\alpha}{\delta E^\varphi} \label{PBeq7} \\
\frac{\delta \tilde{Q}^e}{\delta \eta}\frac{\delta \tilde{\omega}_x}{\delta P^\eta}-
\frac{\delta \tilde{Q}^e}{\delta P^\eta}\frac{\delta \tilde{\omega}_x}{\delta \eta}&=-\frac{1}{2\gamma}\left(2s\frac{\delta \tilde{Q}^e}{\delta E^x}+\frac{E^\varphi}{2|E^x|}\frac{\delta \tilde{Q}^e}{\delta E^\varphi}+p 2\sqrt{2}|E^x|^\frac{1}{4}\frac{\delta \tilde{\omega}_x}{\delta E^\varphi}\right) \label{PBeq8} \\
\frac{\delta Q^\alpha}{\delta \eta}\frac{\delta \tilde{\omega}_x}{\delta P^\eta}-
\frac{\delta Q^\alpha}{\delta P^\eta}\frac{\delta \tilde{\omega}_x}{\delta \eta}&=-\frac{1}{2\gamma}\left(2s\frac{\delta Q^\alpha}{\delta E^x}+\frac{E^\varphi}{2|E^x|}\frac{\delta Q^\alpha}{\delta E^\varphi}\right) \label{PBeq9} \\
\frac{\delta \alpha}{\delta \eta}\frac{\delta \tilde{\omega}_x}{\delta P^\eta}-\frac{\delta \alpha}{\delta P^\eta}\frac{\delta \tilde{\omega}_x}{\delta \eta}&=-\frac{1}{2\gamma}\left(2s\frac{\delta \alpha}{\delta E^x}+\frac{E^\varphi}{2|E^x|}\frac{\delta \alpha}{\delta E^\varphi}\right)\,. \label{PBeq10} 
\end{align}

To try to find a solution to this system of equations we assume
$\tilde{Q}^e$ to be independent of $P^\eta$ and $\eta$. Then
(\ref{PBeq6}) and (\ref{PBeq7}) imply that $\alpha$ and $Q^\alpha$ are
independent of $E^\varphi$, and we are left with equation
(\ref{PBeq5}) and
\begin{gather}
2s\frac{\delta \tilde{Q}^e}{\delta E^x}+\frac{E^\varphi}{2|E^x|}\frac{\delta \tilde{Q}^e}{\delta E^\varphi}+p2\sqrt{2}|E^x|^\frac{1}{4}\frac{\delta \tilde{\omega}_x}{\delta E^\varphi} = 0 \notag \\
\frac{\delta Q^\alpha}{\delta \eta}\frac{\delta \tilde{\omega}_x}{\delta P^\eta}-
\frac{\delta Q^\alpha}{\delta P^\eta}\frac{\delta \tilde{\omega}_x}{\delta \eta}=-\frac{s}{\gamma}\frac{\delta Q^\alpha}{\delta E^x} \notag \\
\frac{\delta \alpha}{\delta \eta}\frac{\delta \tilde{\omega}_x}{\delta P^\eta}-
\frac{\delta \alpha}{\delta P^\eta}\frac{\delta \tilde{\omega}_x}{\delta \eta}=-\frac{s}{\gamma}\frac{\delta \alpha}{\delta E^x}\,. \notag 
\end{gather}
Using (\ref{PBSpinConntilde}) which implies $\frac{\delta \tilde{\omega}_x}{\delta E^\varphi}=0$, the equation for $\tilde{Q}^e$ is generically solved for $\tilde{Q}^e[E^x,\Ef]=h(\Ef/|E^x\,^{\frac{1}{4}})$ with $h$ a function of one variable.
If we further require that $\G^3$ reproduce the Gauss constraint
(\ref{Gauss}), $\G^3=k G_{\rm grav}$ for an arbitrary constant $k$, we find the
solution giving rise to the canonical transformation (\ref{SOLUTION}).

\section{Dilaton gravity with torsion}  \label{altApproach}

For completeness, we summarize here the constructions necessary in the
presence of torsion. This will not change the main results of the
paper.

\subsection{Generalized Dilaton Gravity and comparison of metrics}

Using the definitions~\cite{DilatonRev}
\[
\phi:=\Phi^2 \quad,\quad
g_{\mu\nu}:=\mathbf{g}_{\mu\nu}\quad,\quad
\mathbf{U}(\phi):=-\frac{1}{2\phi}\quad,\quad
\mathbf{V}(\phi):=1
\]
 instead of (\ref{ConformalDefs}), the spherically symmetric reduced
 2d action (\ref{SSreducedAction}) is reexpressed as the generalized
 2d Dilaton action
\[
S_{\rm dilaton}=\frac{1}{2G}\int d^2x\sqrt{-g}\left(\frac{1}{2}\phi R + W(-(\nabla\phi)^2,\phi)\right)\,.
\]
This in turn is equivalent to a general 2d gravity action with
torsion:
\[
S=-\frac{1}{2G}\int_M \phi d\omega -W(X^aX_a,\phi)\varepsilon + X_aDe^a
\]
with
\[
W(X^aX_a,\phi):=\mathbf{U}(\phi)\frac{X^aX_a}{2}+\mathbf{V}(\phi)
\]
and $R=2*d\bar{\omega}$ the curvature for the torsion free part of the
spin connection $\omega=\bar{\omega}+e_a\frac{\partial W}{\partial
X_a}$. There is no conformal transformation, so the metric $g$
represents the physical metric in this approach.

%\footnotetext{This gives an overall factor of $-\frac{1}{2G}$ in the PSM action as opposed to $-\frac{2}{G}$ from before.}

The Poisson sigma model in this case is determined by the more
general Poisson bivector
\[
   \mathcal{P}^{ij}=\begin{pmatrix}
                            0     &   W   &  -X^-  \\
	   	           -W     &   0   &   X^+  \\  
		            X^-   & -X^+  &   0
    	            \end{pmatrix}
\]
so that in the constraints (\ref{Gmp}), $V/2$ is replaced by $-W$:
\begin{equation} \label{GmpT}
\tG^\mp=\frac{1}{2G}\left[\left(\frac{e Q^e\pm Q^\alpha}{2e}\right)'\mp\left(\frac{e Q^e\pm Q^\alpha}{2e}\right)(\omega_x+\alpha')\pm W(2X^+X^-,\phi)e\right]\exp (\mp\alpha)  
\end{equation}
and from (\ref{Xpm0})
\[
2X^+X^-=2\left(\frac{e Q^e - Q^\alpha}{2e}\right)\left(\frac{e Q^e + Q^\alpha}{2e}\right)\,.
\]

Comparison of the metrics now yields different values for $\phi$ and
the dyads (\ref{diads1}) which get an extra factor of
$|E^x|^{-\frac{1}{4}}$:
\begin{align}
  \phi&=|E^x| \notag \\
  e^+_x&=p\frac{E^\varphi}{\sqrt{2}|E^x|^\frac{1}{2}}\exp\alpha\quad,\quad
  e^-_x=p\frac{E^\varphi}{\sqrt{2}|E^x|^\frac{1}{2}}\exp(-\alpha)   \notag \\
  e^+_t&=p\frac{N^xE^\varphi\pm N|E^x|^\frac{1}{2}}{\sqrt{2}|E^x|^\frac{1}{2}}\exp\alpha \quad,\quad
  e^-_t=p\frac{1}{\sqrt{2}|E^x|^\frac{1}{2}}\left(\frac{-N^2|E^x|+N^x\,^2E^\varphi\,^2}{N^xE^\varphi\pm N|E^x|^\frac{1}{2}}\right)\exp(-\alpha) \notag
\end{align}

The dependence of the Lagrange multipliers $X^\pm$ in terms of
$(E^x,E^\varphi,K_x,K_\varphi)$ (as in equations (\ref{Xm},\ref{Xp}))
consequently gets an extra factor of $|E^x|^{\frac{1}{4}}$:
\begin{align}
X^-&=p\sqrt{2}|E^x|^\frac{1}{2}\left(-s\frac{E^x\,'}{2E^\varphi}\mp K_\varphi\right)\exp(-\alpha)   \label{Xm2}\\
X^+&=p\sqrt{2}|E^x|^\frac{1}{2}\left(s\frac{E^x\,'}{2E^\varphi}\mp K_\varphi\right)\exp(\alpha)   \label{Xp2}
\end{align}
The torsion free part of the spin connection $\omega_x$ is
$\bar{\omega}_x=\pm K_x-\alpha'$, and the torsion dependent part
$e_{xa}\frac{\partial W}{\partial X_a}=\mathbf{U}(\phi)e_{xa}X^a=\pm
2\mathbf{U}(E^x)E^\varphi K_\varphi$, so
\begin{equation}
\omega_x=\pm s K_x\mp 2\mathbf{U}(E^x)E^\varphi K_\varphi-\alpha' \label{spinConn22}
\end{equation}

\subsection{Canonical transformation}
The Poisson bracket relations
\[
\{Q^e(x),e(y)\}=\{Q^\alpha(x),\alpha(y)\}=\{\phi(x),\omega_x(y)\}=2G\delta(x,y)
\]
give the following functional dependence of
$(Q^e,Q^\alpha,\phi;e,\alpha,\omega_x)$ in terms of
$(E^x,E^\varphi,P^\eta;K_x,K_\varphi,\eta)$:
\begin{align}
    \phi=|E^x|              \quad&,\quad
       e=\frac{E^\varphi}{\sqrt{2}|E^x|^\frac{1}{2}}                   \notag \\
\omega_x=-s\frac{A_x}{\gamma}-\frac{E^\varphi}{|E^x|}K_\varphi+\tilde{\omega}_x[E^x,E^\varphi,P^\eta,\eta] \quad&,\quad
     Q^e=p\sqrt{2}|E^x|^\frac{1}{2}K_\varphi+\tilde{Q}^e[E^x,E^\varphi,P^\eta,\eta]                              \notag \\
Q^\alpha=Q^\alpha[E^x,E^\varphi,P^\eta,\eta]  \quad&,\quad
  \alpha=\alpha[E^x,E^\varphi,P^\eta,\eta]         \label{FUNCFORM2} 
\end{align}
and the following differential equations analogous to (\ref{PBeq5}),
(\ref{PBeq6}), (\ref{PBeq7}), (\ref{PBeq8}), (\ref{PBeq9}),
(\ref{PBeq10}):
\begin{align}
\frac{\delta Q^\alpha}{\delta \eta}\frac{\delta \alpha}{\delta P^\eta}-
\frac{\delta Q^\alpha}{\delta P^\eta}\frac{\delta \alpha}{\delta \eta}&=\frac{1}{\gamma} \label{PBeq52} \\
\frac{\delta \tilde{Q}^e}{\delta \eta}\frac{\delta \alpha}{\delta P^\eta}-
\frac{\delta \tilde{Q}^e}{\delta P^\eta}\frac{\delta \alpha}{\delta \eta}&=-p\frac{\sqrt{2}|E^x|^\frac{1}{2}}{\gamma}\frac{\delta \alpha}{\delta E^\varphi} \label{PBeq62} \\
\frac{\delta \tilde{Q}^e}{\delta \eta}\frac{\delta Q^\alpha}{\delta P^\eta}-
\frac{\delta \tilde{Q}^e}{\delta P^\eta}\frac{\delta Q^\alpha}{\delta \eta}&=-p\frac{\sqrt{2}|E^x|^\frac{1}{2}}{\gamma}\frac{\delta Q^\alpha}{\delta E^\varphi} \label{PBeq72} \\
\frac{\delta \tilde{Q}^e}{\delta \eta}\frac{\delta \tilde{\omega}_x}{\delta P^\eta}-
\frac{\delta \tilde{Q}^e}{\delta P^\eta}\frac{\delta \tilde{\omega}_x}{\delta \eta}&=-\frac{1}{2\gamma}\left(2s\frac{\delta \tilde{Q}^e}{\delta E^x}+\frac{E^\varphi}{|E^x|}\frac{\delta \tilde{Q}^e}{\delta E^\varphi}+p2\sqrt{2}|E^x|^\frac{1}{2}\frac{\delta \tilde{\omega}_x}{\delta E^\varphi}\right) \label{PBeq82} \\
\frac{\delta Q^\alpha}{\delta \eta}\frac{\delta \tilde{\omega}_x}{\delta P^\eta}-
\frac{\delta Q^\alpha}{\delta P^\eta}\frac{\delta \tilde{\omega}_x}{\delta \eta}&=-\frac{1}{2\gamma}\left(2s\frac{\delta Q^\alpha}{\delta E^x}+\frac{E^\varphi}{|E^x|}\frac{\delta Q^\alpha}{\delta E^\varphi}\right) \label{PBeq92} \\
\frac{\delta \alpha}{\delta \eta}\frac{\delta \tilde{\omega}_x}{\delta P^\eta}-
\frac{\delta \alpha}{\delta P^\eta}\frac{\delta \tilde{\omega}_x}{\delta \eta}&=-\frac{1}{2\gamma}\left(2s\frac{\delta \alpha}{\delta E^x}+\frac{E^\varphi}{|E^x|}\frac{\delta \alpha}{\delta E^\varphi}\right) \label{PBeq102} 
\end{align}

Already, from the form of $\omega_x$ in (\ref{FUNCFORM2}) and in
(\ref{spinConn22}) we see that a canonical transformation is only
consistent for $\mathbf{U}(\phi)=-\frac{1}{2\phi}$. So we only have
freedom to change the functional form of $\mathbf{V}(\phi)$ if we try
to generalize to other models (just as in the conformal approach).

Attempting to solve these differential equations by assuming
$\tilde{Q}^e=0$ and requiring $\G^3$ to reproduce the Gauss constraint
as before gives the transformation
\begin{align}
     Q^e=p2\sqrt{2}|E^x|^\frac{1}{2}K_\varphi +h[|E^x|^{-\frac{1}{2}}E^\varphi]     \quad&,\quad
       e=p\frac{E^\varphi}{\sqrt{2}|E^x|^\frac{1}{2}}                              \notag \\    
    \phi=|E^x|            \quad&,\quad
\omega_x=-sK_x-\frac{E^\varphi}{|E^x|} K_\varphi +\frac{1}{k}\eta'     \notag \\
Q^\alpha=\frac{k}{\gamma}P^\eta+\left(\frac{k-s\gamma}{\gamma}\right)E^x\,' 
\quad&,\quad
  \alpha=-\frac{1}{k}\eta         \label{SOLUTION2} 
\end{align}
and its inverse
\begin{align}
E^x=s\phi             \quad&,\quad
E^\varphi=p\sqrt{2}\phi^\frac{1}{2}e           \notag \\
K_x=-s(\omega_x+\alpha'+\frac{e}{2\phi}(Q^e-h)) \quad&,\quad
K_\varphi=p\frac{(Q^e-h)}{2\sqrt{2}\phi^\frac{1}{2}} \notag \\
\eta=-k \alpha        \quad&,\quad
P^\eta=\frac{\gamma}{k}Q^\alpha+\left(\frac{\gamma-sk}{k}\right)\phi'   \label{INVSOLUTION2}
\end{align}
where again, $s={\rm sign}(E^x)$, $k$ is an arbitrary constant, and
$h$ an arbitrary function of one variable.

\subsection{Constraints}
Again, for $h=0$ we have
\[
\G^3[\lambda]=kG_{\rm grav}[\lambda] \quad,\quad
D[N^x]=D_{\rm grav}[N^x]
\]
and the remaining linear combination is
\begin{align}
\G^+[N\exp(-\alpha)]-\G^-[N\exp(\alpha)]&=\frac{1}{2G}\int dx\,N\left(-Q^e(\omega_x+\alpha')-\left(\frac{Q^\alpha}{e}\right)'-2We\right) \notag\\
=\frac{p\sqrt{2}}{2G}\int dx\,N \bigg[&-|E^x|^{-\frac{1}{2}}K_\varphi^2 E^\varphi - 2s |E^x|^{\frac{1}{2}}K_x K_\varphi -|E^x|^{-\frac{1}{2}} E^\varphi \mathbf{V} \notag \\ 
                     &+\frac{|E^x|^{-\frac{1}{2}} E^x\,'^2}{4E^\varphi}-\frac{s |E^x|^{\frac{1}{2}}E^x\,' E^\varphi\,'}{E^\varphi\,^2}+\frac{s |E^x|^{\frac{1}{2}} E^x\,''}{E^\varphi} \bigg]  \notag \\
+\frac{k}{2G\gamma}\int dx\,N\bigg[&
\frac{sE^x\,'}{2|E^x|^\frac{1}{2}E^\varphi}(E^x\,'+ P^\eta)-\frac{k}{4\gamma|E^x|^\frac{1}{2}E^\varphi}(E^x\,'+ P^\eta)^2 \notag \\
&-\bigg(\frac{|E^x|^{\frac{1}{2}}}{E^\varphi}(E^x\,'+ P^\eta)\bigg)' \bigg]\,. \notag 
\end{align}
For $\mathbf{V}(\phi)=1$, this gives
\begin{align}
\G^+[N\exp(-\alpha)]-&\G^-[N\exp(\alpha)]=p\sqrt{2}H_{\rm grav}[N]  \notag \\
+\frac{k}{2G\gamma}\int dx\,N\bigg[&
\frac{sE^x\,'}{2|E^x|^\frac{1}{2}E^\varphi}(E^x\,'+ P^\eta)-\frac{k}{4\gamma|E^x|^\frac{1}{2}E^\varphi}(E^x\,'+ P^\eta)^2 \notag \\
&-\bigg(\frac{|E^x|^{\frac{1}{2}}}{E^\varphi}(E^x\,'+ P^\eta)\bigg)' \bigg] \notag 
\end{align}

In summary,
\begin{align}
\tG^3&=k\tilde{\mathcal{G}}_{\rm grav}  \label{G3GgT} \\
e\,\exp(\alpha)\tG^-+e\,\exp(-\alpha)\tG^++\omega_x\tG^3&=\tilde{\mathcal{D}}_{\rm grav} \label{DDgT} \\
-\exp(\alpha)\tG^-+\exp(-\alpha)\tG^+-\frac{\phi'-G\tG^3}{2e\phi}\tG^3+\left(\frac{\tG^3}{e}\right)'&=p\sqrt{2}\tilde{\mathcal{H}}_{\rm grav}\,. \label{GmpHgT}
\end{align}

\end{appendix}

%\bibliographystyle{../../CQG/CQG}
%\bibliographystyle{../../preprint}
%\bibliography{../../Bib/QuantGra}

\end{document}